\UseRawInputEncoding
\documentclass[aps,pra,reprint, superscriptaddress,showkeys]{revtex4-2}
\usepackage{caption}
\usepackage[utf8]{inputenc}
\usepackage{amsmath, amssymb, amsthm, mathtools, bbm, physics}
\usepackage{amsmath}
\usepackage{amsthm}
\usepackage{graphicx}
\usepackage{amssymb}
\usepackage{dsfont}
\usepackage{array}

\usepackage{color}
\usepackage{braket}
\usepackage{natbib}
\usepackage{siunitx}
\usepackage{ragged2e}
\usepackage{mathtools}
\usepackage{natbib}
\usepackage[compatibility=false]{caption}
\usepackage{subcaption}
\usepackage{xspace}
\usepackage{tikz}
\usetikzlibrary{arrows.meta}
\usetikzlibrary{positioning}
\usetikzlibrary{calc} 
\usepackage{hyperref}
\usepackage{theoremref}

\definecolor{mycitecolor}{RGB}{150, 120, 200}  
\definecolor{mylinkcolor}{RGB}{0, 102, 51}    
\definecolor{myurlcolor}{RGB}{204, 102, 0}
\hypersetup{
    colorlinks=true,
    citecolor=mycitecolor,
    linkcolor=mylinkcolor,
    urlcolor=myurlcolor
}
\definecolor{darkorange}{rgb}{1, 0.55, 0.0}

\definecolor{olivegreen}{rgb}{0,0.5,0.1}

\theoremstyle{plain}

\AtBeginDocument{\RenewCommandCopy\qty\SI}
\begin{document}
\title{Detection Efficiency Bounds in (Semi-)Device-Independent Scenarios}

\author{Tailan S. Sarubi\href{https://orcid.org/0009-0009-2046-6346}{\includegraphics[scale=0.05]{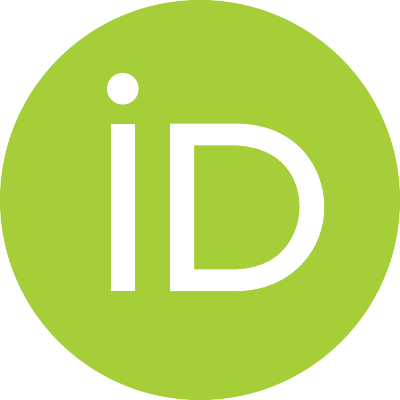}}}
\email{Corresponding author: sarubi.santos@gmail.com}
\affiliation{Physics Department, Federal University of Rio Grande do Norte, Natal, 59072-970, Rio Grande do Norte, Brazil}
\affiliation{International Institute of Physics, Federal University of Rio Grande do Norte, 59078-970, Natal, Brazil}

\author{Santiago Zamora}
\email{fsantiagoz1994@gmail.com}
\affiliation{Physics Department, Federal University of Rio Grande do Norte, Natal, 59072-970, Rio Grande do Norte, Brazil}
\affiliation{International Institute of Physics, Federal University of Rio Grande do Norte, 59078-970, Natal, Brazil}

\author{Moisés Alves}
\email{moisesabc.alves@gmail.com}
\affiliation{Physics Department, Federal University of Rio Grande do Norte, Natal, 59072-970, Rio Grande do Norte, Brazil}
\affiliation{International Institute of Physics, Federal University of Rio Grande do Norte, 59078-970, Natal, Brazil}

\author{Vinícius F. Alves}
\email{vinicius.alves.701@ufrn.edu.br}
\affiliation{Physics Department, Federal University of Rio Grande do Norte, Natal, 59072-970, Rio Grande do Norte, Brazil}
\affiliation{International Institute of Physics, Federal University of Rio Grande do Norte, 59078-970, Natal, Brazil}

\author{Gandhimohan M. Viswanathan\href{https://orcid.org/0000-0002-2301-5593}{\includegraphics[scale=0.05]{orcidid.pdf}}}
\email{gandhi.viswanathan@gmail.com}
\affiliation{Physics Department, Federal University of Rio Grande do Norte, Natal, 59072-970, Rio Grande do Norte, Brazil}

\author{Rafael Chaves\href{https://orcid.org/0000-0001-8493-4019}{\includegraphics[scale=0.05]{orcidid.pdf}}}
\email{rafael.chaves@ufrn.br}
\affiliation{International Institute of Physics, Federal University of Rio Grande do Norte, 59078-970, Natal, Brazil}
\affiliation{School of Science and Technology, Federal University of Rio Grande do Norte, Natal, Brazil}

\begin{abstract}
This article provides a comprehensive review of the critical role of detection efficiency in demonstrating non-classicality across various device-independent and semi-device-independent scenarios. The central focus is the \emph{detection loophole}, a challenge in which imperfect detectors can allow classical hidden variable models to mimic quantum correlations, thus masking genuine non-classicality. As a review, the article revisits the paradigmatic Bell scenario, detailing the efficiency requirements for the CHSH inequality—such as the $2/3$ threshold for symmetric efficiencies—and traces the historical trajectory toward the first loophole-free tests. The analysis extends to other causal structures to explore how efficiency requirements are affected in different contexts. These include: the instrumental scenario, which for binary variables has recently been shown to follow the same inefficiency bounds as the bipartite dichotomic Bell scenario; the prepare-and-measure scenario, where inefficiencies impact the certification of a quantum system's dimension and create security breaches in protocols such as Quantum Key Distribution (QKD); and the bilocality scenario, which exemplifies how employing multiple independent sources can significantly relax the required efficiencies to certify non-classical correlations.
\end{abstract}


\date{\today}

\maketitle

\tableofcontents
\newpage
\section{Introduction}
\label{sec:INTR}

Einstein, Podolsky, and Rosen’s fundamental objection to the completeness of quantum mechanics \cite{einstein1935can} and John Bell’s subsequent mathematical formalization inaugurated one of the most profound areas of modern physics \cite{bell1964einstein}. Bell’s inequalities provided an experimentally testable criterion to distinguish the predictions of quantum mechanics from local realism \cite{Clauser1969,Clauser1972,Aspect1981,Aspect1982}, the classical worldview in which properties exist independently of measurement and no influence can propagate faster than light \cite{bohm1951quantum}. The violation of these inequalities, confirmed in numerous experiments \cite{Giustina2013,Hensen2015,Giustina2015,Shalm2015,rosenfeld2017event,BigBellTest2018}, not only validated the nonclassical nature of quantum entanglement but also laid the groundwork for revolutionary quantum technologies such as quantum cryptography and quantum computing \cite{nielsen2010quantum,Ekert91,GHOREISHI20251}.

However, the transition from theory to experimental practice revealed significant challenges, the most critical of which is the \emph{detection loophole} \cite{Pearle}. This loophole arises because no detector is perfectly efficient; particle losses are inevitable and can be exploited by a hidden-variable model to mimic quantum correlations \cite{Branciard2011,Eberhard93}, thus masking genuine nonclassicality. Closing this loophole, by ensuring that detection efficiency exceeds a critical threshold, is the key to an irrefutable demonstration of nonclassicality and to certifying the security of quantum protocols in a device-independent manner \cite{Ekert91}, in which one does not rely on the internal workings of the measurement devices.

Beyond the canonical Bell scenario, the challenge of detection efficiency remains critical in other causal structures used to test nonclassicality \cite{Pearl2009causality}. For instance, the instrumental scenario \cite{pearl2013,chaves2018quantum,van2019quantum,GMR_QCI} assesses nonclassicality under different causal premises, leading to distinct efficiency requirements. In semi-device-independent contexts, such as the prepare-and-measure scenario \cite{gallego2010}, detector inefficiency can directly compromise the certification of a quantum system’s dimension~\cite{DallArno2015,Mironowicz2021}. Furthermore, quantum networks, exemplified by the paradigmatic bilocal scenario \cite{branciardbilocal,branciardcharacterizing,tavakoli2021bilocal}, exploit correlations from multiple independent sources. This network topology imposes unique constraints on classical models, which, in turn, alter the efficiency thresholds required to observe nonclassical effects. Analyzing these varied requirements is therefore essential for developing robust nonclassicality tests tailored to diverse experimental platforms.

This article offers a unified perspective on the role of detection efficiency across multiple scenarios and includes novel results within the instrumental scenario. It is organized as follows: in Sec.~\ref{sec:The Bell scenario} we revisit the Bell scenario, detailing the efficiency requirements for the CHSH inequality and its multipartite generalizations, such as the Mermin and Svetlichny inequalities. Sec.~\ref{sec:The Instrumental Scenario} explores the instrumental scenario, an asymmetric causal scenario with communication, and analyzes how the combination of observational and interventional data affects efficiency thresholds. Sec.~\ref{sec:The Prepare-and-Measure Scenario} addresses inefficiency thresholds in the prepare-and-measure scenario. In Sec.~\ref{sec:The Bilocality Scenario} reviews the bilocality scenario, showing how a network topology with independent sources can relax the efficiency demands for certifying non-classicality. Finally, in Sec. \ref{sec:final} we provide an overview of the topic and discuss interesting directions for future research.

\section{The Bell Scenario}
\label{sec:The Bell scenario}

The concept of a Bell scenario traces its roots back to the 1930s, when Einstein, Podolsky, and Rosen (EPR) formulated their famous argument challenging the completeness of quantum mechanics \cite{einstein1935can}. At that time, the foundations of the theory were under intense debate, and the EPR paper highlighted the tension between quantum predictions and classical notions of reality. To properly understand what constitutes a Bell scenario, we need first to revisit the EPR argument.

The EPR argument considers a maximally entangled state of two particles, which are distributed to  space-like separated laboratories $A$ and $B$. When an observer in laboratory $A$ measures the position or momentum of their particle, entanglement allows for the prediction of the measurement results of the particle in laboratory $B$. This is regardless of the distance between $A$ and $B$. Under these circumstances, EPR argue that the position and momentum of the particle in B are both ``elements of reality," as it is possible to predict their values with certainty and without disturbing the system in $B$, simply by performing local measurements in $A$. On the other hand, quantum theory does not allow for the same state to assign definite values to both the position and momentum of a particle simultaneously; therefore, EPR concluded that the quantum state cannot be a complete description of physical reality.

This argument was later reformulated by David Bohm in a simpler setting, involving two spin-1/2 particles in the singlet state \cite{bohm1951quantum},
\begin{equation}\label{bellstates}
    \ket{\Psi^-}_{AB} = \frac{1}{\sqrt{2}}(\ket{01}-\ket{10}),
\end{equation}
where $\ket{0}$ and $\ket{1}$ are the eigenstates of the spin operator in the $z$-direction, $S_z = \frac{\hbar}{2}(\ket{0}\bra{0}-\ket{1}\bra{1})$, whose eigenvalues are $+\hbar/2$ and $-\hbar/2$, respectively. In this state, measurements of the spin components of A and B in any direction will have perfectly anticorrelated results. Thus, if $A$ measures the spin of their particle in the \(z\)-direction and obtains $+\hbar/2$ ($-\hbar/2$), then $B$ will obtain the result $-\hbar/2$ ($+\hbar/2$) if they perform the same measurement on their part of the quantum system. Similarly, if $A$ and $B$ measure the spin in the \(x\)-direction, they will also always obtain opposite results. 

Therefore, by choosing to measure the spin of their particle in the \( z \)- or \( x \)-direction, $A$ can infer the corresponding measurement results of these components for the particle at $B$. In this case, one can consider that orthogonal components of the spin of a particle, such as \(z\) and \(x\), are elements of reality. Thus, they should both have a counterpart in any complete description of the physical system. However, in quantum mechanics, these quantities are described by operators that do not commute, so if one has complete knowledge about one of the spin components of a particle, then the theory says nothing about the other orthogonal components. Hence, if the premises of this argument are accepted, one may conclude that the quantum state does not provide a complete description of the physical reality.

In 1964, motivated by the EPR argument, John Bell proved a crucial result that became known as Bell's theorem \cite{bell1964einstein}, which asserts that quantum theory is not compatible with all the premises assumed by EPR. Since these premises are not fulfilled by the theory, the conclusion of the argument, which is that the quantum-mechanical description of reality must be incomplete, does not hold. To formalize the EPR's assumptions, Bell introduced the concept of ``local-realism", which appears in the context of the so-called Bell scenario, where a source sends physical systems to a number of spatially separated laboratories. In the simplest case, which is called a bipartite Bell scenario, one considers two distant laboratories, $A$ and $B$, which receive systems from a common source. In laboratory $A$, an observer has a set of possible measurement choices that can be performed on their system, from which one is chosen; this measurement choice is denoted by $x$. Similarly, the observer in laboratory B chooses to perform a measurement labeled $y$. The measurement outcomes of $A$ and $B$, which we denote by $a$ and $b$, may vary in different runs of the experiment, therefore, they are generally described by the conditional probability distribution $p(a,b|x,y)$, that contains all the information about the correlations between the measurement results of $A$ and $B$.

To explain these correlations, one may invoke a causal assumption called \textit{Reichenbach's principle of common causes}, which states that if two variables are correlated, but do not causally influence each other directly, then there should exist a common-cause variable that influences them both \cite{reichenbach1991direction}. In the context of a Bell scenario, since the laboratories $A$ and $B$ may be arbitrarily far apart, then, according to the principle of locality, the measurement choices and results of $A$ should not causally influence the measurement outcomes of $B$, and vice versa. Therefore, following Reichenbach's principle, there should exist a variable $\lambda$ which explains the correlations between the measurement results $a$ and $b$. Since $\lambda$ is not directly part of the accessible data of the experiment, it is called a latent, or hidden variable, and the assumption of its existence is referred to as ``realism".

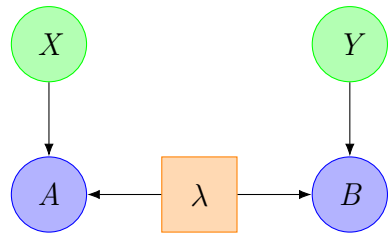
\begin{figure}[h!]
\centering
\begin{tikzpicture}[
    node distance=1.5cm,
    ab circle node/.style={circle, draw=blue, fill=blue!30, minimum size=1cm, inner sep=0pt, font=\large\itshape}, 
    other circle node/.style={circle, draw=green, fill=green!30, minimum size=1cm, inner sep=0pt, font=\large\itshape}, 
    every rectangle node/.style={rectangle, draw=orange, fill=orange!30, minimum size=1cm, inner sep=0pt, font=\large\bfseries}, 
    ->, >={Latex}
]

\node[ab circle node] (A) at (0,0) {A};
\node[ab circle node] (B) [right of=A, xshift=2.5cm] {B};
\node[every rectangle node] (Lambda) at ($(A)!0.5!(B)+(0,0)$) {$\lambda$};
\node[other circle node] (X) [above of=A, yshift=0.5cm] {X};
\node[other circle node] (Y) [above of=B, yshift=0.5cm] {Y};

\draw[->] (X) -- (A);             
\draw[->] (Y) -- (B);             
\draw[->] (Lambda) -- (A);       
\draw[->] (Lambda) -- (B);       
\end{tikzpicture}
\caption{\justifying \textbf{Directed acyclic graph representing the causal assumptions of local-realism and measurement independence in the bipartite Bell scenario.}}
\label{fig:localcausal}
\end{figure}

The premises of local-realism may be expressed by means of a \textit{directed acyclic graph} (DAG), which is a way of specifying the assumed causal relations between a set of variables \cite{Pearl2009causality, Wood_2015, chaves2015}. In a DAG, each node represents a variable, and the arrows represent causal influence relationships between them. This is shown in Fig.~\ref{fig:localcausal}, where, for instance, the arrow going from $X$ to $A$ means that the measurement result $a$ is conditionally dependent on the measurement choice $x$. In this DAG, locality is represented by the absence of arrows going from one party's measurement choice or measurement result to the other party's measurement result. For instance, neither the choice $x$ nor the outcome $a$ of party $A$ should influence the outcome $b$ of party $B$. Moreover, the assumption of realism is represented by the hidden variable $\lambda$, and the arrows going from \(\lambda\) to $a$ and $b$ represent its causal influence on the measurement results of both parties. For completeness, we mention that there is an extra assumption implicit in the DAG: the measurement independence assumption \cite{hall2010local,barrett2011much,hall2011relaxed,putz2014arbitrarily,chaves2015,chaves2021causal,vieira2025test,BigBellTest2018,Handsteiner2017,scheidl2010violation}. The absence of an arrow between the input $X$ and $Y$ and $\lambda$ implies that the parties are free to independently choose their measurement. 

The assumptions of local-realism imply certain restrictions on the possible correlations that can be attained between the measurement results of $A$ and $B$. Probabilistic models that are compatible with these hypotheses are called local hidden variable models (LHV), and their form is given by the following expression:
\begin{equation}
    p(a,b|x,y) = \int_{\Lambda} d\lambda\, q(\lambda) p(a|x,\lambda)p(b| y,\lambda),
    \label{LHV model}
\end{equation}
where the probability density $q(\lambda)$ of the hidden variable reflects the fact that it may vary across different runs of the experiment, as it is not necessarily a controllable parameter.

In contrast to local-realism, in quantum theory, the probabilities are given by the Born rule, which has the following form in the case of a bipartite Bell scenario:
\begin{equation}
    p(a,b|x,y) = \mathrm{Tr}\left[(M^A_{a|x}\otimes M^B_{b|y}) \rho_{AB}\right].
    \label{born rule bell}
\end{equation}
In expression (\ref{born rule bell}), $M^A_{a|x}$ and $M^B_{b|y}$ are measurement operators of $A$ and $B$, generally described by Positive Operator Valued Measures (POVMs), and $\rho_{AB}$ is the density operator describing the state of the system. The approach for proving that local hidden variable models cannot reproduce the predictions of quantum theory is based on comparing the correlations allowed by the Born rule (\ref{born rule bell}) to the ones constrained by the causal assumptions given by (\ref{LHV model}).

The simplest possible nontrivial Bell scenario for doing that is the one in which there are only 2 measurement choices and 2 possible results for each of the 2 parties, which can be labeled $x,y,a,b\in\{0,1\}$. In 1969, Clauser, Horne, Shimony and Holt showed that any local hidden variable model in this scenario obeys the so-called CHSH inequality \cite{Clauser1969}:
\begin{equation}
\abs{S}\equiv\abs{\expval{A_0 B_0} + \expval{A_0B_1} + \expval{A_1 B_0} - \expval{A_1 B_1}}\leq 2,
    \label{CHSH inequality}
\end{equation}
with the expectation values $\expval{A_xB_y}$ defined by $\expval{A_xB_y} = \sum_{a,b\in\{0,1\}}(-1)^{a+b}p(a,b|x,y)$. The incompatibility of quantum theory with local-realism, which became known as quantum nonlocality, can thus be proved by showing that quantum systems can violate inequality (\ref{CHSH inequality}). In fact, if $A$ and $B$ share any bipartite pure entangled state, the CHSH inequality can be violated with suitable local measurements, a result known as Gisin's theorem \cite{gisin1991bells}, that shows the essential role of entanglement for quantum nonlocality (although the essential role of entanglement has been questioned in a recent work \cite{wang2025violation}). Furthermore, values of $\abs{S}$ as large as $\abs{S^{max}_{QM}} = 2\sqrt{2}$ can be attained quantum-mechanically \cite{cirel1980quantum}, a result that lies well beyond the predictive scope of LHV models.

\subsection{Revisiting the Journey to Loophole-Free Bell Tests}
\label{historic}

As mentioned above, in principle, a Bell test experiment can distinguish quantum theory from any LHV model. However, experimental loopholes \cite{larsson2014} have historically hindered definitive conclusions. These loopholes arise from deviations from ideal assumptions, allowing for local-realist explanations for the observed results. The three more significant are the locality detection efficiency, and freedom of choice loopholes, even though other forms of loopholes have also been identified \cite{barrett2002quantum,bendersky2016algorithmic}. The locality loophole arises if the measurements on the two sides are not truly simultaneous or not sufficiently space-like separated, thus permitting light-speed signals to connect the devices and invalidating the assumption of no causal communication between Alice and Bob. The detection loophole arises when a significant fraction of measurement events go undetected, which can occur due to imperfect detector efficiency. To analyze such experiments as if all events were recorded, researchers often invoke the fair-sampling assumption \cite{berry2010fair}, which posits that the subset of detected events is representative of the entire ensemble. However, if this assumption does not hold, a local-realist model could exploit the detection inefficiencies to selectively conceal outcomes that would otherwise violate local-realism, thereby creating a false agreement with quantum predictions \cite{Branciard2011}. Finally, the freedom-of-choice loophole concerns the independence of measurement settings from any hidden variables, necessitating that settings be genuinely random and free from external influences.

Focusing on the detection efficiency loophole, a local-realistic model can exploit detection inefficiencies by deliberately failing to produce outcomes (i.e., causing lost events) precisely in those cases where it would otherwise violate a Bell inequality. As a result, it can yield a biased subset of detected events that appears to violate the inequality while still being fully compatible with local-realism. Indeed, Branciard \cite{Branciard2011} has demonstrated that one can construct post-selected local correlations, conditioned only on events where both detectors click, that violate the CHSH inequality in a manner that mimics genuine nonlocality, even though they originate from a local-realistic model exploiting the detection loophole. These observations underscore the importance of closing all experimental loopholes; otherwise, a skeptic can always attribute the observed violations to imperfections in the setup rather than to the intrinsic nonclassicality of quantum theory.

The vulnerability of the detection loophole explains why early Bell tests, despite being historic, were not conclusive. Freedman and Clauser's 1972 experiment \cite{Clauser1972} tested the Clauser-Horne (CH) inequality \cite{Clauser.Horne} using calcium's $J=0 \rightarrow 1 \rightarrow 0$ cascade, which emitted photons at wavelengths of 551.3\,nm and 422.7\,nm. Utilizing pile-of-plates polarizers with transmittances $e_\parallel \approx 0.97$, $e_\perp \approx 0.038$  and inefficient detectors (efficiencies around $\eta_1 \approx 1.7 \times 10^{-3}$, $\eta_2 \approx 1.5 \times 10^{-3}$), they relied on the fair-sampling assumption. After 200 hours of data collection, they observed a 6 standard deviation violation of the normalized $CH$ parameter, $\delta = 0.050 \pm 0.008$, marking the first Bell inequality violation; however, both the locality and the detection loopholes remained unaddressed.

Nearly a decade later, Aspect, Grangier, and Roger (1981) \cite{Aspect1981} refined the test by employing two-photon laser excitation (406.7\,nm and 561\,nm) in the same calcium cascade, 
producing $4 \times 10^7$ photon pairs per second within a compact $60\,\mu\text{m} \times 1\,\text{mm}$ interaction volume. With “pile-of-plates” polarizers characterized by $e_{\parallel} = 0.971 \pm 0.005$, $e_{\perp} = 0.029 \pm 0.005$, and visibility $F = 0.984$, they measured 150 true coincidences/s (within a window 19\,ns). For angular settings $(0^\circ, 22.5^\circ, 45^\circ, 67.5^\circ)$, they reported $S^{CH}_{\text{exp}} = 0.126 \pm 0.014$, violating Bell’s bound by 9 standard deviations, in excellent agreement with the quantum prediction $S^{CH}_{\text{QM}} = 0.118 \pm 0.005$. The Freedman-style $\delta$-violation reached 13 standard deviations.

In 1982, Aspect, Grangier, and Roger performed two landmark experiments \cite{Aspect1982b, Aspect1982}.
In the first \cite{Aspect1982b}, they implemented two-channel analyzers using dichroic cubes ($T_{\parallel} \approx 0.95$, $R_{\perp} \approx 0.93$) with cross-talk below 1\%, enabling simultaneous acquisition of the four coincidence rates $R_{ij}(a,b)$.
This allowed direct calculation of the correlation function $E(a,b)$, resulting in $S = 2.697 \pm 0.015$, a 46 standard deviation violation of Bell's inequality, reaching 83\% of the quantum limit $2\sqrt{2} \approx 2.828$ \cite{cirel1980quantum}.
The second experiment \cite{Aspect1982} used acousto-optic modulators to randomly switch settings during the photon flight time ($\sim 43\,\text{ns}$, switch time $\sim 10\,\mu\text{s}$), reporting $S = 2.701 \pm 0.015$. Although aiming to address the locality loophole (without closing the detection efficiency issue), the global deterministic control signal meant the setting choices were not truly independent, leaving the proof of non-classicality only partially loophole-free.

In the late 1980s, more precisely in 1988, Z.~Y.~Ou and L.~Mandel reported the first violation using Type I parametric down-conversion (PDC) in potassium dihydrogen phosphate (KDP) \cite{Mandel88}. Unlike complex atomic sources, this configuration produced quantum interference via post-selection, with a coincidence probability given by $P(\theta_1, \theta_2) = K \sin^2(\theta_1 - \theta_2)$, which results in a CHSH violation of 6 standard deviations. Despite detection efficiencies $\eta \ll 1$, the experiment established a stable and scalable optical protocol, although both the locality and detection loopholes remained open.

Seven years later, in 1995, P.~G.~Kwiat et al.\  developed a non-collinear Type-II PDC source in a BBO crystal capable of directly generating photon pairs in all four Bell states, enabling the production of maximally entangled two-qubit states without the need for post-selection~\cite{Kwiat1995}. These so-called Bell states form an orthonormal basis for the Hilbert space of two qubits and are given by:
\begin{equation}
\begin{array}{r@{\quad}r}
|\Phi^\pm\rangle = \frac{1}{\sqrt{2}} (|00\rangle \pm |11\rangle), &
|\Psi^\pm\rangle = \frac{1}{\sqrt{2}} (|01\rangle \pm |10\rangle).
\end{array}
\end{equation}
With a measured visibility of $(97.8 \pm 1.0)\%$ and count rates exceeding 1500 s$^{-1}$, they achieved a CHSH violation greater than 100 standard deviations in just 5 minutes. Detection efficiencies of approximately $\sim10\%$ and operational robustness made this source a benchmark for subsequent Bell tests.

 A significant milestone was reached by Weihs et al. (1998), who decisively closed the locality loophole using spatially separated measurement stations, placed 400\,m apart \cite{Weihs1998}. The relativistic communication time ($\sim 1.3\,\mu$ s) exceeded the duration of each measurement sequence ($<100$, ns), which incorporates physical randomness of LED-based quantum random number generators ($t_c \lesssim 10$\,fs), electro-optical switching ($\tau_\text{delay} = 75$\,ns), and detection. Assuming fair-sampling (with detection efficiency $\eta \approx 5\%$), they observed a strong violation of the CHSH inequality with $S = 2.73 \pm 0.02$, corresponding to a 30 standard deviations deviation from the local bound. However, the detection loophole remained unaddressed.

To directly address the detection loophole, Rowe et al. (2001) employed trapped $^9\text{Be}^+$ ions as entangled qubits \cite{Rowe2001}. They prepared the Bell state $|\Psi\rangle = \frac{1}{\sqrt{2}} (|\uparrow\uparrow\rangle - |\downarrow\downarrow\rangle)$ with a fidelity of approximately $0.88$, and achieved near-unity detection efficiency ($\eta > 98\%$) via state-dependent fluorescence. The resulting Bell parameter, $S = 2.25 \pm 0.03$, violated the CHSH inequality by 8 standard deviations without requiring auxiliary assumptions.  Nevertheless, the close proximity of the ions (3\,$\mu$m) left the locality loophole open.

Still in 2001, a proposed technology marked a paradigm shift in the detection of Bell tests. Goltsman et al. introduced the first superconducting nanowire single-photon detectors (SNSPDs), demonstrating that the absorption of a single photon in an ultrathin NbN nanowire creates a localized resistive hotspot detectable in nanoseconds \cite{Goltsman2001}. In the following years, key optimizations were achieved: between 2007 and 2009, meander-style geometries and improvements in impedance matching increased internal detection efficiency to approximately 50--70\% \cite{Kerman2007, Anant2008}; from 2012 to 2014, Marsili et al. reached a greater efficiency of 90\% at telecom wavelengths (1550\,nm), with timing jitter below 20\,ps \cite{Marsili2013}. These advances established SNSPDs as ideal candidates to close the detection loophole and to enable spatio-temporal conditions sufficient for a genuinely loophole-free Bell test \cite{Christensen2013}.

In 2015, Hensen et al. conducted the first Bell test simultaneously free of detection and locality loopholes \cite{Hensen2015}. The experiment used electron spins in nitrogen–vacancy (NV) centers in diamond, separated by 1.3~km. Entanglement was generated via entanglement swapping, correlating each spin with a photon and performing a joint measurement at an intermediate station. A fidelity of approximately 92\% was achieved. Measurement settings were chosen by low-latency quantum random number generators, with spacetime separation enforced between setting choice and measurement. Spin detection was performed by resonant fluorescence with near-unit efficiency, eliminating the fair-sampling assumption. Over 245 events, they obtained a CHSH parameter of $S = 2.42 \pm 0.20$, violating the classical bound by about two standard deviations.

That same year, the NIST group in Boulder implemented a loophole-free Bell test using polarization-entangled photons produced via Type-II parametric downconversion in a PPKTP crystal \cite{Shalm2015}. Photons were directed to stations 180~m apart, with bases selected by ultrafast random number generators combining optical phase diffusion and prestored pseudorandom data, ensuring independence from the source. Detection was performed using single photon superconducting nanowire detectors (SNSPDs) with system efficiencies greater than 74\%, closing the detection loophole. Statistical analysis accounting for non-IID effects and 0.3\% residual input predictability yielded an adjusted $p$-value of $2.3 \times 10^{-7}$, exceeding 6 standard deviations violation of the Clauser–Horne inequality. Spacetime separation among events ensured the simultaneous closure of locality and detection loopholes.

In parallel, the Vienna group led by Zeilinger conducted a loophole-free Bell test using a Sagnac interferometer configuration \cite{Giustina2015}. A pulsed 405\,nm diode laser pumped a PPKTP crystal, generating polarization-entangled photons directed to measurement stations separated by 58\,m. Measurement bases were chosen via electro-optic modulators driven by quantum random number generators based on spontaneous emission. Detection was performed using high-efficiency transition edge sensors (TES), achieving heralding efficiencies of 78.6\% and 76.2\%. The test evaluated the CH-Eberhard parameter, obtaining $J = 7.27 \times 10^{-6}$ over 3510 seconds, with a corresponding value $p$ of approximately $3.74 \times 10^{-31}$, equivalent to a violation of 11.5 standard deviations. The experiment successfully closed both the locality and detection loopholes, yielding one of the most compelling photon-based violations of Bell’s hypothesis.

Two milestone experiments addressed the freedom-of-choice loophole by ensuring that measurement settings were independent of any shared causal past with the entangled source. In 2017, Handsteiner \textit{et al.} used light from high-redshift quasars emitted more than 7 billion years ago to determine the analyzer settings in a Bell test over 180~m, obtaining $S = 2.65 \pm 0.02$ \cite{Handsteiner2017}. This ruled out hidden variable models exploiting correlations within the entire past light cone of the experiment. In 2018, the Big Bell Test \cite{BigBellTest2018} distributed random bits generated by over 100,000 human participants to 12 labs across five continents, achieving Bell violations in all platforms and closing the freedom-of-choice loophole under conservative assumptions of biased yet independent human randomness.

The journey toward loophole-free Bell tests represented a multi-decade quest to validate the most counter-intuitive predictions of quantum mechanics. Continuing with the initial experiments in the 1970s, which despite being groundbreaking, were compromised by locality and detection loopholes, the scientific community progressively refined its techniques. Subsequent experiments improved particle sources, measurement speed, and detector efficiency. The culmination of this journey was the execution of tests that simultaneously closed the principal loopholes, furnishing robust proof against local-realism. This monumental effort was ultimately recognized with the 2022 Nobel Prize in Physics, awarded to Alain Aspect, John Clauser, and Anton Zeilinger for their pioneering experimental contributions that transformed the understanding of the nature of quantum entanglement.

By closing the locality loophole and constraining freedom of choice, any viable local-realist model seeking to account for the observed correlations must posit that its hidden variables were fixed on cosmological timescales, before the quasar photons used in the experiments were even emitted. This requirement, that the causal history of those variables predates nearly the entire observable universe, renders such explanations virtually implausible; nevertheless, a superdeterminist could still argue that these correlations originate at the very beginning of the universe. Real-time setting choices based on astronomical or human sources ensure space-like separation between the measurement choices and outcomes. However, high detection efficiency remains crucial to eliminate post-selection bias. As demonstrated by Eberhard, a reliable CHSH violation ($S > 2$) is only possible when detection efficiency exceeds a critical threshold $\eta_{\mathrm{min}}$ \cite{Eberhard93}. Therefore, combining strict space-like separation with high detection efficiency is essential for a loophole-free Bell test, leaving local-realism without any credible refuge.

\subsection{Modeling of Imperfect Detectors}\label{sec:ND_models}

In this section we present two general models of non-ideal detectors and how this affects the observed statistics based on \cite{Wilms2008}. Let us start by defining exactly what we mean by a detector.  In the context of Bell-type experiments, a detector refers to the physical or effective measurement device associated with a party that registers an outcome upon receiving a particle or quantum system. Formally, it maps an input setting $x$ to a classical output $a$, producing a conditional probability distribution $p(a|x)$. A detector is said to fire when it produces a valid outcome, and fails when it produces no result (e.g., a no-click event). The detection efficiency $\eta$ quantifies the probability that the detector fires when it should, playing a central role in closing the detection loophole.

In practice, various factors such as optical losses or  sensor limitations, prevent the detection of 100\% of all events. The simplest way in which we can model an imperfect  detector is by assuming that on each measurement attempt, the detector  registers a valid outcome with probability $\eta$ and fails (no click) with 
probability $1-\eta$. Mathematically, if $p(a|x)$ is the probability of obtaining the outcome $a$ given the measurement $x$ in an ideal scenario (with perfect detection), then in the real scenario with efficiency $\eta$ we have
\begin{equation}
    p_{\eta}(a|x) = \eta p(a|x),
\end{equation}
for each conclusive outcome $a$. Consequently, by representing the absence of outcome  by the symbol $\varnothing$, then the probability of no detection given the measurement $x$ will be
\begin{equation}
    p_{\eta}(\varnothing|x) = (1 - \eta) \sum_a p(a|x) =1-\eta.
\end{equation}

That is, the detector fails with probability $1 - \eta$, independently of the ideal outcome distribution (assumption of outcome-independent detection). This is the simplest independent misdetection model, frequently assumed in theoretical analysis of the detection loophole \cite{Garbarino2010}. 

We can extend this to $n$ detectors. To illustrate let us take $n=2$, for Alice and Bob, with efficiencies 
$\eta_1$ and $\eta_2$. If both perform joint measurements  each round, four 
situations can occur: both detectors fire, only Alice's detector fires, only Bob's detector fires, or neither detector fires. In the independent model, 
the probabilities of these occurrences factor as $\eta_1 \eta_2$, $\eta_1 (1 - \eta_2)$, $\eta_2 (1 - \eta_1)$, and $(1 - \eta_1)(1 - \eta_2)$, respectively \cite{Garbarino2010}. For example, the  coincidence  probability (the probability when both detectors click) will be
\begin{equation}
    p_{\eta_1 \eta_2}(a,b|x,y) = \eta_1 \eta_2 p(a,b|x,y),\label{eq: Model_abs_1}
\end{equation}
whereas the probability that only Alice's detector clicks with outcome $a$ but Bob does not register anything  is~\cite{Garbarino2010, Wilms2008}
\begin{equation}
    p_{\eta_1 \eta_2}(a,\varnothing|x,y) = \eta_1 (1 - \eta_2)\sum_b p(a,b|x,y).\label{eq: Model_abs_3}
\end{equation}
An analogous equation can be written when Bob's detector clicks and Alice's does not, and the case where both detectors do not click can be found simply by satisfying the normalization constraint of the probabilities.  From now on we will call this model the \textit{extra-outcome model} or simply \textit{Model 1}.

Another kind of model introduced in Ref.~\cite{Wilms2008} is when instead of considering an extra output $\varnothing$, one absorbs the no-click events in an output $\hat o$ for example.  This could be useful for example when modeling a polarization detector: in principle the no-click event can be due to a photon with the wrong polarization or in fact because the detector did not fire as it should. Then we can absorb this missed click in the wrong polarization output. In the case of two detectors we can write 
\begin{equation}
    p_{\eta_1 \eta_2}(a,b|x,y) = \eta_1\eta_2 p(a,b|x,y),\label{eq:abs_1}
\end{equation}
whenever we actually detect outcomes $a$ and $b$. Now let us assume that we cannot distinguish between the outcome $\hat b$ and the no-click event. Formally we can write
\begin{eqnarray}
    p_{\eta_1 \eta_2}(a,\hat b| x, y) =& \eta_1\eta_2 p(a,\hat b|x,y) + \nonumber \\& \eta_1(1-\eta_2)\sum_b p(a,b|x,y).\label{eq:abs_2}
\end{eqnarray}

Last expression considers the probability of actually obtaining the outcome $\hat b$ plus the probability of not having detected anything in $B$.  As in model 1, an analogous equation can be written for Alice's side and the expression representing to the no-click  outcome can be obtained using the normalization of $p_{\eta_1 \eta_2}$. In the following sections we will call this \textit{absorption model} or \textit{model 2}.


Models of this kind preserve the no-signaling structure of the data (Alice’s no-detection cannot depend on Bob’s choice $y$, for example). In such a case then we have $\sum_b p(a,b|x,y) = p(a|x)$ for example, where $ p(a|x)$ is the marginal on Alice's side. Furthermore, these models allow rigorous analysis of which post-selected probability distributions (conditioned on double click events for example) admit a local explanation \cite{Wilms2008}.

\subsection{Efficiency Requirements in the CHSH Scenario}

Now we restrict to some of the theoretical results regarding the critical efficiencies needed for detection loophole-free proofs of nonlocality in the Bell scenario. As reviewed  in Sec.~\ref{historic}, in experimental Bell tests it is common that not all pairs of emitted particles are detected. As demonstrated by Pearle \cite{Pearle}, ignoring events where a detector registered no signal (``no click'') allows local hidden variable models to reproduce quantum correlations, even in scenarios where the Bell inequality is violated in the detected sample. 

A genuine violation of local-realism through a Bell inequality in post-selected scenarios requires the observed correlations to lie outside the post-selected LHV polytope ($L_{ps}$) \cite{Branciard2011}. The facets of $L_{ps}$ define the valid Bell inequalities for this context. In the CHSH scenario, this process generates a modified Clauser-Horne \cite{Clauser.Horne} inequality denoted by $I_{CH}^{\eta_1,\eta_2} \leq 0$. Analysis in Ref.~\cite{Branciard2011} reveals that violation of this inequality by any non-signaling theory requires satisfaction of a fundamental necessary condition on detection efficiencies:
\begin{equation}
\eta_1 + \eta_2 < 3\eta_1\eta_2,
\end{equation}
If this condition is unmet, the $L_{ps}$ polytope expands to coincide with the entire non-signaling polytope $\mathcal{P}_{ns}$ \cite{BrunnerREVIEW}, rendering Bell violations impossible \cite{Branciard2011}. This condition is therefore the gateway to detection loophole-free tests.

For a bipartite dichotomic Bell test with symmetric efficiencies ($\eta_1 = \eta_2 = \eta$), the general condition reduces to $2\eta < 3\eta^2$. Solving for $\eta > 0$ yields the celebrated Eberhard bound \cite{Eberhard93}:
\begin{equation}
\eta > \frac{2}{3} \approx 66.7\%,
\end{equation}
This represents the minimum efficiency required to close the detection loophole in a CHSH test using quantum states. Interestingly, this bound is achieved with non-maximally entangled states, demonstrating that maximum-violation correlations are not necessarily the most loss tolerant. This contrasts sharply with the more stringent requirement of $\eta > 2(\sqrt{2} - 1) \approx 82.8\%$ for states with maximal entanglement \cite{Clauser.Horne}. 

Significant improvements in efficiency thresholds emerge when considering asymmetric detection efficiencies ($\eta_1 \neq \eta_2$). The canonical example involves pairs of entangled atom-photons, where atomic states are measured with near-perfect efficiency ($\eta_1 \approx 1$), while photon detection is typically inefficient ($\eta_2 < 1$). Applying the general condition $3\eta_1\eta_2 - \eta_1 - \eta_2 > 0$ yields the Larsson-Cabello bound for asymmetric efficiencies \cite{CabelloLarsson}. For the experimentally critical case of one perfect detector ($\eta_1 = 1$), the condition simplifies to:
\begin{equation}
\eta_2 > \frac{1}{2} = 50\%,
\end{equation}
This shows a substantial reduction in the efficiency threshold for the lossy party (from 66.7\% to 50\%), enabling new loophole-free experimental configurations \cite{Teo2013}.

A deeper analysis considers efficiencies that depend on the specific
observable being measured (e.g., $\eta_1$ for one setting $A_1$ and
$\eta_2$ for a setting $B_1$). Using a Hardy-type argument,
Garbarino introduces a family of nonmaximally entangled two-qubit
states $|\psi_H(\alpha)\rangle$, parametrized by a real overlap
parameter $0<\alpha<1$ that relates the eigenbases of $\hat a$ and
$\hat b$ \cite{Garbarino2010}. For this family, the probability of the
Hardy event $P_{\rm H}(\alpha)=P(a^+,a^+)$ is
$P_{\rm H}(\alpha) = (1-\alpha^2)^2 \alpha^2 / (2-\alpha^2)$, and one
can show that a loophole-free test is possible whenever
$\eta_1 > 2(1-\eta_2)/\alpha^2$. In particular, if one observable
($B_1$) can be measured with perfect efficiency ($\eta_2 = 1$), then
in the ideal noiseless case the Bell inequality can be violated for
any nonzero efficiency of the other observable ($\eta_1>0$).

A complementary approach was developed by Cabello, Larsson, and Rodriguez, who analyzed the minimum detection efficiencies required to violate the Braunstein-Caves (BC) chain Bell inequalities without invoking the fair-sampling assumption \cite{Cabello2009}. In contrast to Garbarino's asymmetric, observable-specific condition, their framework derives tight detection bounds for both symmetric (\(\eta_1 = \eta_2\)) and asymmetric scenarios. For the symmetric case, they establish that a loophole-free violation of the BC inequality with \(N\) measurement settings per party is possible if and only if the detection efficiency exceeds \(\eta_{\text{crit}} = 2 / \left[N \cos\left(\pi / 2N\right) + 1\right]\). In the asymmetric case, where one party detects with unit efficiency, the other must satisfy \(\eta_1 > (N-1)/[N \cos(\pi/2N)]\). Their analysis provides a clear operational link between detection efficiency and the algebraic structure of the inequality, revealing that higher-setting BC inequalities, while useful for other quantum information tasks, demand increasingly stringent efficiencies as \(N\) grows.

The Eberhard, Larsson-Cabello and Garbarino bounds were initially derived through distinct methods. An elegant unification presented by Quintino in Ref.~\cite{quintino2012black} incorporates efficiencies $\eta_{1}^i$ and $\eta_{2}^j$ depending on both the party and the measurement setting ($i,j$). Bell violation occurs if and only if at least one of four symmetric conditions is satisfied, such as:

\begin{align}\nonumber
\eta_{1}^0\eta_{2}^0 + \eta_{1}^0\eta_{2}^1 + \eta_{1}^1\eta_{2}^0 - \eta_{1}^0 - \eta_{2}^0 > 0, \\ \nonumber
\eta_{1}^0\eta_{2}^1 + \eta_{1}^0\eta_{2}^0 + \eta_{1}^1\eta_{2}^1 - \eta_{1}^0 - \eta_{2}^1 > 0, \\\nonumber
\eta_{1}^1\eta_{2}^0 + \eta_{1}^1\eta_{2}^1 + \eta_{1}^0\eta_{2}^0 - \eta_{1}^1 - \eta_{2}^0 > 0, \\
\eta_{1}^1\eta_{2}^1 + \eta_{1}^1\eta_{2}^0 + \eta_{1}^0\eta_{2}^1 - \eta_{1}^1 - \eta_{2}^1 > 0, \\\nonumber
\end{align}
Necessity follows from Branciard's polytope analysis \cite{Branciard2011}, while sufficiency is proven via explicit construction of violating quantum states and measurements. All previous bounds emerge as corollaries under appropriate efficiency definitions \cite{quintino2012black}.

The CHSH inequality is not the only tool for combating the detection loophole. Brunner \textit{et al.} analyzed inequality $I_{3322}$ with three measurement settings per party \cite{BrunnerI3322}. In an asymmetric setup with one perfect detector ($\eta_1 = 1$), combined with weakly entangled states, this reduces the threshold for the second detector to:
\begin{equation}
\eta_2 \gtrsim 43\%,
\end{equation}
This highlights how choosing the Bell inequality combined with quantum state optimization can dramatically reduce experimental efficiency demands, enabling robust demonstrations of quantum non-classicality under substantial detection losses.

\subsubsection{Numerical examples}

\begin{figure}[!ht]
            \centering           
            \includegraphics[scale=0.55]{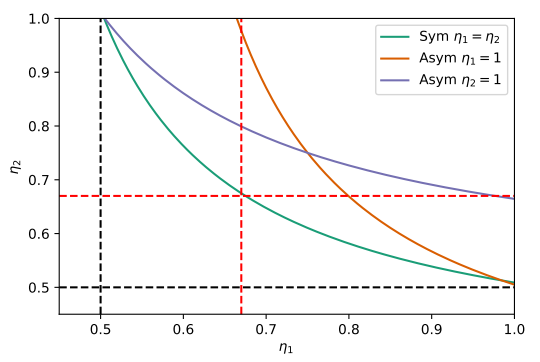}
            \caption{ \justifying \textbf{Violation of the CHSH absorbing the non-detection events in \(\hat{a}=1\) and \(\hat{b}=0\). }In the case where \(\hat{a} = 1\) and \(\hat{b} = 0\), the figure depicts the boundary curve \(\eta_2(\eta_1)\) beyond which a violation of the CHSH inequality is no longer possible. For the symmetric configuration, represented by the green curve, the critical efficiency threshold is \(\eta_1 = \eta_2 = 0.67\), corresponding to the intersection point of the two red dashed lines. The green curve itself was obtained with the optimized measurements and states that give the critical efficiencies \(\eta_1 = \eta_2 = 0.67\). In the asymmetric regime with one ideal detector (\(\eta_1 = 1\)), the minimum efficiency for the other detector is \(\eta_2 = 0.50\); by symmetry, the same holds if \(\eta_2 = 1\) and \(\eta_1 = 0.50\). The purple and orange curves are obtained by using the  optimized configurations for \(\eta_1 = 0.51\) and \(\eta_2 = 0.51\), respectively.}
            \label{fig2}
    \end{figure}

\begin{figure}[!ht]
            \centering           
            \includegraphics[scale=0.55]{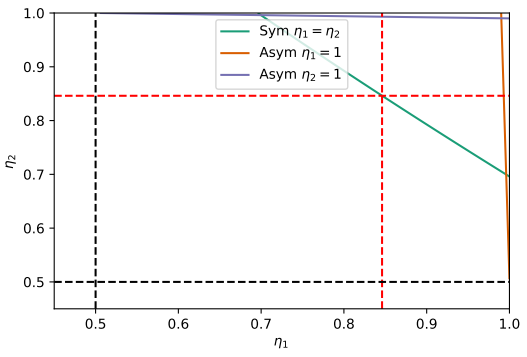}
            \caption{ \justifying \textbf{Violation of the CHSH absorbing the non-detection events in \(\hat{a}=1\) and \(\hat{b}=1\)}. In the figure we plot the curve \(\eta_2(\eta_1)\) below which no violation of the CHSH inequality is possible when no-detection events are absorbed in \(\hat a = 1\) and \(\hat b = 1\). In the symmetric case (\(\eta_1 = \eta_2\), green curve), the minimum efficiency required is \(\eta_1 = \eta_2 = 0.84\), identified by the intersection of the two red dashed curves. In the asymmetric case, with one perfect detector (\(\eta_1 = 1\)), the minimum occurs at \(\eta_2 = 0.50\), and analogously, if \(\eta_2 = 1\) is fixed then the minimum over \(\eta_1\) is also \(0.50\). The orange and purple curves indicate, respectively, the optimized boundary conditions for \(\eta_1 = 0.51\) (fixing \(\eta_2\)) and for \(\eta_2 = 0.51\) (fixing \(\eta_1\)). The curves were obtained with the optimal measurements and state that give the minimum critical efficiency in the symetric and asymmetric case.  }
            \label{fig1}
    \end{figure}

We complete this subsection by illustrating the above analytical bounds
with explicit numerical optimizations. In all cases, we fix a specific
loss model, parametrize the underlying two-qubit state and local
projective measurements, and, for each pair of efficiencies
$(\eta_1,\eta_2)$, we maximize the corresponding Bell functional. The boundary curves $\eta_2(\eta_1)$ shown in Fig.~\ref{fig2} and~(\ref{fig1}) are defined
as the locus of points for which the maximal quantum value of the Bell
expression coincides with its local bound. For efficiencies above this
curve a violation is possible ($I_{\max} > \beta_{\mathrm{loc}}$), whereas
below it no violation can be obtained.

The first model of inefficient detectors that we consider arises
naturally in the context of quantum optics. In a Bell-type experiment,
Alice and Bob must distinguish between two orthogonal polarizations
(e.g.\ horizontal/vertical). However, we assume that each party
possesses only a single on/off detector sensitive to one of the
polarizations. Consequently, a ``no-click'' event is indistinguishable
from the arrival of a photon with the wrong polarization: both are
registered as the complementary outcome of the dichotomic observable
(denoted by $\hat a$ or $\hat b$). Operationally, this corresponds to
an absorption model in which the ideal projective measurement is
preceded by a lossy channel of efficiency $\eta_j$ on each side $j=1,2$,
and the no-detection events are absorbed into one of the two outcomes.
In the optimizations that follow we implement this model explicitly in
the probabilities used to evaluate the Bell expressions.

This ambiguity leads to a modified relationship between the ideal Bell probabilities \(p^{\text{Bell}}(a,b | x,y)\) and the noisy probabilities observed with detection efficiencies \(\eta_1\) and \(\eta_2\), denoted \(p^{\text{Bell}}_{\eta_1 \eta_2}(a,b | x,y)\). These non-ideal probabilities are given by the absorption model described in Sec.\ref{sec:ND_models} (See also~\cite{Wilms2008}).

Using Eq.~(\ref{eq:abs_1}) and Eq.~(\ref{eq:abs_2}), the corresponding expression for Alice's detector and the normalization condition,    the CHSH inequality~(\ref{CHSH inequality}) can be optimized for both the symmetric detection case (\(\eta_1 = \eta_2\)) and the maximally asymmetric case (\(\eta_1 = 1\) and optimization over \(\eta_2\)). For the CHSH, one obtains the known thresholds reported in Ref.~\cite{Wilms2008}: in the symmetric case, the minimum efficiency is \(\eta_m^s \approx 0.6667\), while in the asymmetric case, we find \(\eta_m^{as} \approx 0.5003\).The numerical results depend critically on how no-detection events are absorbed. We will illustrate two cases:  a scenario that recovers the well-known $~67\%$ threshold (Fig.~\ref{fig2}), and second, a different scenario that leads to a much stricter requirement of $84\%$ (Fig.~\ref{fig1}). Figures~\ref{fig2} and \ref{fig1} show the critical curves above which a loophole-free violation of the CHSH inequality can be achieved. The curves are obtained with the measurements and states that optimize the efficiencies in the symmetric case ($\eta_1 = \eta_2 \approx 2/3$) and in the asymmetric case ($\eta_1 =1, \eta_2 =0.50003$ and vice versa): once the measurements and the state are fixed, the only parameters left are $\eta_1$ and $\eta_2$.

As an alternative modeling approach one can consider non-detection events as additional measurement outcomes rather than discarding them, this is the extra-outcome model introduced in \ref{sec:ND_models}. This perspective naturally broadens the set of possible outcomes, resulting in a higher-dimensional probability framework. Within this structure, both ideal and inefficient scenarios can be uniformly addressed through probabilistic modeling, such as Eqs. (\ref{eq: Model_abs_1}) and (\ref{eq: Model_abs_3}) (plus the analogous equation for Alice and the normalization condition).  By considering the Eberhard inequality~\cite{Eberhard93}, where $e.g.$ outcomes $0,1$ are clicks, $2$ is the no-click event, and settings are $x,y \in \{0,1\}$,


\begin{equation}
\begin{aligned}
E_b =\ & p(01 | 01) + p(02 | 01) + p(10 | 10) + p(20 | 10) \\
       & + p(00 | 11) - p(00 | 00) \geq 0,
\end{aligned}
\end{equation}
 one can  obtain the efficiencies for this type of modeling. The numerical thresholds obtained from this optimization are identical to those found for the CHSH inequality under the absorption model shown in Fig.~\ref{fig2}. 




Finally, we consider the absorption model  with higher dimensional systems, in particular with qutrits.
 The quantum state considered is
\begin{equation}
|\psi\rangle = \gamma_1 |00\rangle + \gamma_2 |11\rangle + \gamma_3 |22\rangle,
\end{equation}
with amplitude coefficients parametrized as $(\gamma_1, \gamma_2, \gamma_3) = (\cos \varphi_0 \sin \theta_0, \sin \varphi_0 \sin \theta_0, \cos \theta_0)$, enabling exploration of both maximally and non-maximally entangled regimes.

To characterize operational behavior under local measurements, we follow the framework of high-dimensional quantum protocols~\cite{Collins2002}, in which each party applies a phase gate with tunable phases $e^{i\varphi_a(j)}$ and $e^{i\varphi_b(j)}$, followed by a discrete Fourier transform. The joint quantum state then evolves to:

\begin{equation}
|\psi\rangle = \frac{1}{3} \sum_{j,k,l=0}^2 \gamma_j e^{i(\varphi_a(j) + \varphi_b(j) + \frac{2\pi}{3} j(k - l))} |k\rangle_A |l\rangle_B,
\end{equation}
Subsequently, Alice and Bob measure in the Fourier basis to produce the final joint distribution:
\begin{equation}
P_{QM}(A_a = k, B_b = l) = \frac{1}{9} \left| \sum_{j=0}^2 \gamma_j e^{\frac{2\pi i j}{3} (\alpha_a + \beta_b + k - l)} \right|^2.
\end{equation}
The phases $\alpha$ and $\beta$ correspond to Alice's and Bob's measurements respectively.
This framework allows for a direct evaluation of the extent to which quantum resources can violate generalized Bell-type inequalities such as the Collins-Gisin-Linden-Massar-Popescu (CGLMP) inequality~\cite{Collins2002} 
\begin{align}
    I_3 &= P(A_1=B_1) - P(A_1=B_1-1)  \nonumber \\
    &+ P(B_1=A_2+1) - P(B_1=A_2)  \nonumber\\ 
    &+ P(A_2=B_2) - P(A_2=B_2-1)   \nonumber\\
    &+ P(B_2=A_1) - P(B_2=A_1-1)  \le 2.
\end{align}

Through numerical optimization one can recover the known maximal violation for a non-maximally entangled state~\cite{Acin2002}.

\begin{figure}[!ht]
            \centering           
            \includegraphics[scale=0.50]{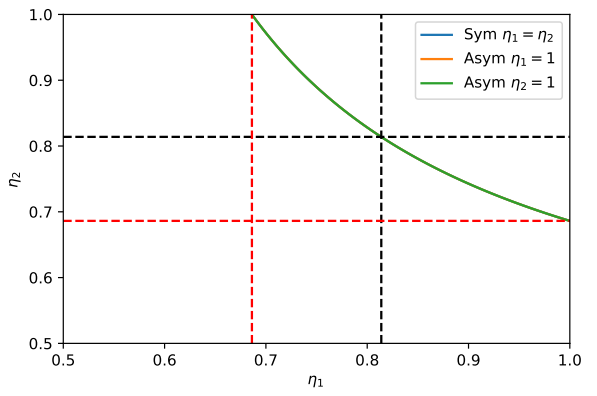}
            \caption{ \justifying  \( I_{3} \) \textbf{inequality with qutrits}. The non-detection event on side $A/B$ is absorbed in the outcome  \( 2 \). This figure shows the values of \( \eta_2(\eta_1) \) for which no violation of the \( I_3 \) inequality occurs. In the symmetric configuration, the minimum efficiency required is \( \eta = 0.81 \). In the asymmetric case, the critical value of \( \eta = 0.68 \) applies regardless of which side is perfect (\( \eta_1 = 1 \) or \( \eta_2 = 1 \)). All curves coincide in the plot. The curves were obtained with the optimal measurements and state that give the minimum critical efficiency in the symetric and asymmetric case. 
}
            \label{fig10}
    \end{figure}

We then incorporate detection inefficiencies using the formalism of the absorption model, where non-detection events are effectively absorbed into a determined outcome. By absorbing the non-detection event in the outcome $2$ in both parties, we determined that a minimum efficiency of $\eta = 0.81$ is required in the symmetric case to close the detection loophole. In the asymmetric regime, where one detector is ideal, the minimal efficiency required by the other detector is $\eta = 0.68$. This behavior is illustrated in Fig.~\ref{fig10}, where all configurations converge to identical thresholds, reflecting a robustness in the qutrit-based configuration.

\subsection{Efficiency Requirements in the Multipartite Scenario}

In this subsection, we review some results in multipartite Bell scenarios.  The study of detection efficiencies in such scenarios is driven by both fundamental and practical interests. On a fundamental level, exploring these scenarios allows for the discovery of novel forms of quantum correlations that only emerge when three or more particles are entangled. Understanding the efficiency requirements to observe these genuinely multipartite effects provides deeper insights into the nature of quantum non-locality and the boundaries between the quantum and classical worlds. Practically, multipartite entanglement is a key resource for advanced quantum information protocols, such as secure quantum communication networks and measurement-based quantum computing. Characterizing the required efficiencies is therefore essential for designing and implementing robust and scalable quantum technologies that can tolerate real-world imperfections like particle loss.

Pinpointing the first seminal paper where efficiencies in multipartite scenarios were studied is difficult, as the analysis evolved through the years. However, we can identify a few related papers that were foundational in the multipartite scenario. First, we mention Ref.~\cite{Svetlichny_ineqs} (1987), where two inequalities were constructed that are satisfied by any bi-separable state, but states with three-body correlations can violate it. These of course, today are called  Svetlichny inequalities (SI) and showcase the richness of non-classicality when considering many-body systems. Another seminal paper is  \textit{Going beyond Bell's theorem} (1989)~\cite{GHZ_Beyond_Bell} , where the authors introduced a multipartite scenario to show that not even in the simplest case where the parties perform measurements in parallel (or anti-parallel) directions of the spin of four entangled particles, and therefore can make definite predictions, can be explained by deterministic local models. This is in contrast to the $2$ party Bell inequality that does not say anything about this case. Inspired by this direct refutation of the EPR argument by means of considering $4$-particle states, which seemed to show stronger correlations than those of  $2$-particle cases, in 1990 Mermin considered $n$ $1/2$-spin particles and proposed a set of $n$-partite inequalities which now carry his name \cite{Mermin_ineqs}. The inequalities as presented by him,  are
\begin{equation}
    -\beta_n \leq \langle\mathcal{B}\rangle\leq \beta_n,
\end{equation}
where $\mathcal{B} \equiv \frac{1}{2i} \left( \prod_{j=1}^{n} \left( \hat{\sigma}_{x}^{j} + i\hat{\sigma}_{y}^{j} \right) - \prod_{j=1}^{n} \left( \hat{\sigma}_{x}^{j} - i\hat{\sigma}_{y}^{j} \right) \right)$, with $\hat \sigma^j_i$ are pauli matrices of the $j$-th party. The classical bound is $\beta_n = 2^{(n-1)/2}$ if $n$ is odd or $\beta_n = 2^{n/2}$ if $n$ is even.

It was shown that quantum theory can exceed the classical bound of a Bell inequality by an exponential amount that grows with $n$ as $O(2^{n/2})$. However, the joint efficiency of $n$ detectors necessarily also declines exponentially. Therefore, it was argued at the time that these multipartite systems were not a suitable arena for loophole-free proofs of nonlocality.  

For completeness, we mention that for the $3$-partite scenario with $2$ dichotomic settings, the Mermin and Svetlichny inequalities are representatives of two distinct classes of the facets of the classical polytope, which is completely characterized by the so-called Sliwa inequalities \cite{SLIWA2003165}. 

Efficiencies in the Mermin inequalities were studied quickly after. In 1993, Braunstein and Mann~\cite{Mermin_eff_93} studied signal-to-noise ratio using them and showed that the minimum efficiency required for a loophole-free violation of the Mermin inequalities in the absence of noise is above $2^{(1-n)/2n}$ for $n$ odd or above  $2^{(2-n)/2n}$ for $n$ even.

Around ten years later, several other notable works on multipartite efficiencies began to emerge. In particular, we mention those of Larsson \cite{PhysRevA.59.4801, CH_n_sites, Larsson_Bell, Larsson_GHZ_1, Larsson_GHZ_2}.
In \cite{Larsson_GHZ_1} the GHZ paradox was analyzed in detail (see Ref.~\cite{hwgb-g93f} for modern discussions on the GHZ paradox).  There,  conditions on the efficiency thresholds without assuming independent errors or requiring knowledge of the particle triple emission rate were derived, which broadens their applicability compared to previous bounds. The GHZ paradox yields a contradiction if at least one of the following efficiency conditions is met: a single-particle efficiency greater than 5/6, a two-particle efficiency greater than $4/5$, or a three-particle efficiency ($h_{3,2}$) greater than $3/4$, along with another three-particle efficiency ($h_{3,1}$) greater than $3/5$. The three-particle efficiency ($h_{3,2}$) is highlighted as the most important bound when adding the last detector, being generally valid and estimable by dividing the triple coincidence rate by the double coincidence rate at two detectors. Conversely, if none of these conditions are satisfied, a local-variable model can be constructed to reproduce the quantum-mechanical statistics, meaning no contradiction arises. The core of the derivation centers on demonstrating that a contradiction occurs if and only if the probability of all six random variables describing the measurement results being defined simultaneously is greater than zero. In the follow-up article \cite{Larsson_GHZ_2}, it is shown that the bounds do not improve even by imposing independent errors and constant efficiency detection.

Another interesting work is from Ref.~\cite{CH_n_sites} where a study of efficiencies was presented by generalizing the CH inequality to multipartite scenarios.  Consider an $n$-partite scenario with $2$ dichotomic settings $X_i$, where $i$ denotes the party, and $X={A,B}$ the performed measurement. Denote any subset of parties by  $S\subseteq\{1,\dots,n\}$. Define also the joint probability $E_S$ as the probability of finding $1$ by all $n$ parties, whenever the parties in $S$ measure $B$ while all the parties not in $S$ measure $A$. Also denote $p_S$ as the joint probability of all parties obtaining the value $1$ when measuring $A$. The $n$-partite CH inequality defined can be written as 

\begin{align}
\sum_{|S|=1} E_S - \sum_{k  \text{ even}} \sum_{|S|=k} E_S \le \sum_{|S|=n-1} p_S - (n-1) p_{\{1, 2, \dots, n\}}.
\end{align}

In Ref.~\cite{CH_n_sites} Larsson and  Semitecolos showed that by considering an equal detection efficiency $\eta$ per party and the independent error and fair-sampling assumption, the above $n$-site CH inequality can be violated if 
\begin{equation}
    \eta \geq\frac{n}{2n-1}, \label{eq: Ef_CH_multi}
\end{equation}
Note that  for $n=2$ one recovers $2/3$ obtained by Eberhard~\cite{Eberhard93}. However, more importantly, see that as soon as $n>2$, the scenario allows for lower detection inefficiencies, showing the power of  multipartite scenarios. Soon after, Massar and Pironio~\cite{Massar_Pironio_2003} showed that this is indeed a tight bound, so the critical efficiency $\eta_{crit}$ for these inequalities is  $\eta_{crit} = n/(2n-1)$.

Some years later,  in 2005, Brassard, Broadbent, and Tapp obtained an upper bound for the Mermin inequalities: the minimum critical efficiency is below $\eta <2^{(2-n)/n}$\cite{A_Broadbent_Mermin}. Then, in 2008 Cabello et al. \cite{Cabello_Mermin} obtained a tight condition on the efficiencies of these inequalities for general $n$. The authors show that for $n$ parties, the Mermin inequalities have a critical efficiency of exactly $\eta = n/(2n-2)$ when considering an n-partite GHZ state.  Note that in the limit of a high number of parties, the critical efficiency coincides with the limit of Eq.~(\ref{eq: Ef_CH_multi}).

Regarding the Svetlichny inequalities, in Ref.~\cite{Svet_ineqs_ineff} the authors derive an alternative form for them, the violation of which directly implies a loophole-free violation of the original SI, even in the presence of detection inefficiencies. For the three-particle SI, assuming equal and independent detection efficiencies for all observers, they calculate a required efficiency $\eta_{crit}$  for a loophole-free violation of approximately $\eta_{crit} \approx 0.9666$ . This is determined by comparing the quantum mechanical maximum value of the generalized SI with the derived inequality's bound, which depends on the detection efficiency. Furthermore, the study extends this analysis to the general $n$-particle SI, providing an analytic expression for $\eta_{crit}$. A notable finding is that as the number of particles increases, the threshold efficiency ($\eta_{crit}$) monotonically and rapidly approaches $1$. This contrasts sharply with for example the Mermin inequalities, where,   the threshold efficiency can decrease with increasing particle number. The authors argue that this difference is due to the exponential dependence on the number of parties of the (increasing) classical bound in the Mermin inequalities: in the SI, the bound is independent of the number of parties. 

Around this same year, a study using continuous variable systems came out~\cite{Chaves_Brask_2011}. The work assesses the feasibility of loophole-free nonlocality tests using multipartite W states, specifically focusing on implementations involving a single photon shared among multiple parties. The authors developed a general error model (POVMs) and utilized Bell inequalities and linear programming (polytope approach) to determine required detection efficiencies.
Their findings show that while the $W$ state's nonlocality increases with the number of parties $n$,  its robustness to amplitude damping decreases for $n> 2$ when using what they called Cabello's inequality~\cite{Cabello_inequality}.  For pure photonic $W$ states, achieving loophole-free tests with tight Bell inequalities requires a single-photon detection (SPD) efficiency of approximately .85 (optimal for $n=4$ or $n=7$ with homodyning, or $n=4$ with displacement). In contrast, for atom-photon entangled $W$ states, the optimal thresholds are much lower at $n=2$, reaching as low as .37 SPD efficiency with perfect atom-photon coupling, though higher $n$ values correspond to more entangled states. The polytope approach confirmed similar thresholds but showed slight improvements for n $\ge$ 5, with $\eta_z \approx .33$ when $\eta_x \approx 1$. Here $\eta_x$ and $\eta_z$ represent the efficiency of measuring the $\sigma_x$ or $\sigma_z$ component of the spin, respectively. When loss events can be detected and treated as a third outcome, locality is violated if $\eta_x > 2(1 - \eta_z)$, meaning any nonzero $\eta_x$ suffices if $\eta_z$ approaches 1. Overall, the paper concludes that while increasing $n$ doesn't always improve loss thresholds (and can sometimes worsen them), the scaling of these thresholds is not severe, making multipartite nonlocality tests feasible with only minor improvements in current detection efficiencies.

Another more recent (2018) interesting work is \cite{Polonia_Tamas_Multipartite}, the authors there construct a multipartite inequality considering the following scenario. Considering $n$ parties, they assume only the last $k$ parties to have efficient detectors, while the first $n-k$ have only limited detection efficiency. Their proposal is to set the first $n-k$ parties to measure only some arbitrary projector $\Pi_i^+$ for $i\in\{1,\dots ,n-k\}$, whereas the  $k$ efficient detection parties can measure randomly two observables $A^j_i = \Pi_{ij}^+ -  \Pi_{ij}^- $ for $i\in\{n-k+1,\dots ,n\}, j\in\{1,2\}$.  Then, assuming one knows some already defined Bell inequality $\braket{I_{n-k+1,\dots ,n}}\leq L$ on the last $k$ parties, their inequality reads
$p_1p_2\dots p_{n-k}\braket{I_{n-k+1,\dots ,n}-L}\leq 0$,
 where $p_i = Tr(\rho\Pi^+_i)$ is the probability of the $i$-th party of obtaining the $+$ result. 

Using this inequality, they show that multipartite states like GHZ and cluster states can violate Bell inequalities with just two highly efficient detectors, while the rest can have arbitrarily low (but non-zero) efficiency. Specifically, using the CHSH inequality on the two efficient detectors, the critical efficiency $\eta_{crit}$ is $\approx 0.8284$ ($2/(1+\sqrt{2})$). This can be further reduced to $2/3$ using the Eberhard inequality. This is especially interesting when considering composed systems (for example, atoms + photons), where one of the subsystems allows for high detection efficiencies while the other does not necessarily. 

Finally, to conclude this section, we briefly review two articles that deal with the detection loophole in genuine multipartite nonlocality (GMN) tests. 

The first results are from  Ref.~\cite{Gebhart_2022_GMN} where they acknowledge the problem of demonstrating GMN. In general, to demonstrate it, post-selection is required to filter only the coincidence detections giving rise to the detection loophole. In their work, by using causal diagrams and $d$-separation rules,  Gebhart and Smerzi show that when having perfect detection efficiencies in the detectors, the post-selection process does not actually create a loophole. Whenever the detectors are not perfect, this is no longer true. To solve this,  they derive sharpened Bell inequalities following Ref.~\cite{Larsson_Bell}. With this approach, they obtain threshold efficiencies for the CHSH, Mermin and Svetlichny inequalities: $0.83$, $0.75$, $0.967$ respectively.

\begin{figure}[h!]
\centering
\begin{tikzpicture}[
    node distance=1.3cm,
    abc circle node/.style={circle, draw=blue, fill=blue!30, minimum size=1cm, inner sep=0pt, font=\large\itshape},
    other circle node/.style={circle, draw=green, fill=green!30, minimum size=1cm, inner sep=0pt, font=\large\itshape},
    every rectangle node/.style={rectangle, draw=orange, fill=orange!30, minimum size=1cm, inner sep=0pt, font=\large\bfseries},
    ->, >=Latex, scale=1.2, thick, shorten >=2pt, shorten <=2pt
]

\node[other circle node] (X) at (0,3.5) {X};
\node[other circle node] (Y) at (2,3.5) {Y};
\node[other circle node] (Z) at (4,3.5) {Z};

\node[abc circle node] (A) at (0,1.8) {A};
\node[abc circle node] (B) at (2,1.8) {B};
\node[abc circle node] (C) at (4,1.8) {C};

\node[every rectangle node] (Lambda) at (2,0) {$\lambda$};

\draw[->] (X) -- (A);
\draw[->] (Y) -- (B);
\draw[->] (Z) -- (C);

\draw[->] (Lambda) -- (A);
\draw[->] (Lambda) -- (B);
\draw[->] (Lambda) -- (C);

\end{tikzpicture}
\caption{\justifying \textbf{DAG depicting the causal structure of a tripartite Bell scenario.}}
\label{fig:tripartiteBell_spaced}
\end{figure}
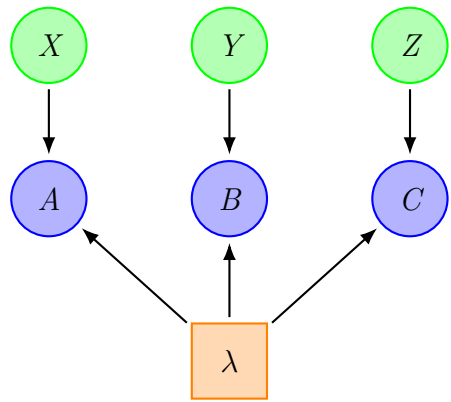

The second article is from Ref.~\cite{efficiencies_Subhendu_2024}. There, the authors calculate the minimum detection efficiency necessary for loophole-free tests of GMN, focusing on both T2-type and Svetlichny-type nonlocality. In particular, they focus on $3$-partite scenarios  (see Fig.~\ref{fig:tripartiteBell_spaced}).  
T2-type locality refers to a definite causal order in the inputs/outputs generation of the parties as a causal constraint. In practice, a 3-party distribution  $p(abc|xyz)$ is said to be T2 local if it can be written as 

\begin{align}
p(abc|xyz) = & \sum_{\lambda} q_{\lambda} p_{\lambda}^{T_{AB}}(ab|xy)p_{\lambda}(c|z) \\
             & + \sum_{\mu} q_{\mu} p_{\mu}^{T_{AC}}(ac|xz)p_{\mu}(b|y) \\
             & + \sum_{\nu} q_{\nu} p_{\nu}^{T_{BC}}(bc|yz)p_{\nu}(a|x),
\end{align}
where $p_{\lambda}^{T_{AB}}(ab|xy)  = p_\lambda^{A<B}(ab|xy)$ if  $B$ is in the future light cone of $A$, or 
$p_{\lambda}^{T_{AB}}(ab|xy)  = p_\lambda^{B<A}(ab|xy)$ if $A$ is in the future light cone of $B$. The effect of this causal constraint is to restrict the possible directions of communication between any two parties \cite{Def_ofNL_2013Bancal, Gallego_NL_2012}.
For T2-type nonlocality, the MDE is found to be 0.75 for each symmetric detector, or 0.5 if two detectors are perfect, exhibiting robustness to noise. For the Svetlichny-type nonlocality, by numerical methods considering all possible tripartite preparations, they show that the minimum efficiency required is of approximately .881 for symmetric detectors, reducing to approximately $.51$ when two detectors are perfect.

For other works including several measurements, or a particular number of parties we refer to the following articles:  in Ref.~\cite{Pal_Vertesi_GHZ},   they consider the GHZ state attaining the lowest efficiency $\eta_{crit} = 12/17$  using $m = 17$ settings per party, improving the bound of Larsson ($0.75$)~\cite{Larsson_GHZ_1, Larsson_GHZ_2}. In Ref.~\cite{Pal_Vertesi_W},  they obtained a lower bound of 0.501 for eight measurement settings by considering $W$ states. In~\cite{M_measurements_1, M_measurements_2, M_measurements_3}, by considering an infinite number of parties, they can find a bound on the efficiency for $M$ measurements of $2/(2+M)$. A work regarding non-signaling theories (2016)  is ~\cite{Cao2016} by Cao and Peng and a work regarding symmetric multiparite Bell inequalities (2024) is~Ref.~\cite{Simmetric_Multipartite_designolle}.

\subsection{Quantum Inefficiencies}
\label{sec: eff_amplitude}

In this section, we review how inefficiencies can arise in the measurements due to losses in the transmitted quantum systems. In general, the discrepancy between the ideal quantum mechanical probabilities and the observed laboratory data can be attributed to loss or noise processes acting between the source and the detector. These effects can be encapsulated by a completely positive trace preservation map (CPTP) $\Phi$, such that the experimentally observed probability for outcome~$j$ is 

\begin{equation*}
    p_j = \mathrm{Tr}\bigl[\Phi_t^{AD}(\rho)\,M_j\bigr],
\end{equation*} 
where $\rho: \mathcal{H} \rightarrow \mathcal{H}$ denotes the prepared quantum state and $\{M_j\}$ is the POVM describing the measurement \cite{nielsen2010quantum}. 

In optical settings where losses correspond to energy dissipation, $\Phi$ reduces to the \emph{amplitude damping} map $\Phi_t^{AD}$. The parameter $0 \le t \le 1$ quantifies the fraction of excitations that survive transit: starting from $n$ photons, the probability of detecting $m$ photons is given by the binomial distribution $P(m | n) = \binom{n}{m}\,t^m\,(1-t)^{\,n-m}$.

In the Kraus operator formalism, the amplitude damping channel is represented as
\begin{equation}\label{AD}
    \Phi_t^{AD}(\rho) = \sum_{k=0}^\infty F_k(t)\,\rho\,F_k^\dagger(t),
\end{equation}
with
\begin{equation*}
    F_k(t)
= \sum_{n=k}^\infty \sqrt{\binom{n}{k}}\;\bigl[t^{\,n-k}(1-t)^{\,k}\bigr]^{\tfrac12}\;\lvert n-k\rangle\langle n\rvert,
\, k=0,1,...,
\end{equation*}
satisfying the completeness relation $\sum_{k=0}^\infty F_k^\dagger(t)\,F_k(t)=\mathbb{I}$. Each operator $F_k(t)$ describes exactly the loss of $k$ photons from the Fock state $\lvert n\rangle$.

For instance, for $n=1$, the channel predicts the transition $\lvert1\rangle \;\to\;\lvert0\rangle
\quad\text{with probability}\quad
1 - t$,
modeling the absorption of single photons into the environment \cite{mandel1995optical}. Photodetection errors originate from imperfect channel transmittance $t$, modeled by the amplitude-damping channel $\Phi_t^{AD}$. In single photon measurement schemes, the total detection efficiency $\eta_{\mathrm{tot}}$ combines channel transmittance $t$ and detector efficiency $\eta$:
\[
\eta_{\mathrm{tot}} = t \,\eta,
\]
where $t$ denotes the probability of a photon surviving transmission and $\eta$ the probability of its subsequent detection. By expressing both effects through the same Kraus operator expansion in Eq. (\ref{AD}), one obtains a cohesive framework for all linear attenuation processes. This unified treatment allows the derivation of a critical total efficiency \(\eta_c\), below which violations of Bell’s inequalities cannot be observed.

In the Heisenberg picture, the losses described by the amplitude-damping channel $\Phi_t^{AD}$ in the Schrödinger picture can be transferred to the measurement operators through its dual map $\Phi_t^{*\,AD}$, defined by
\begin{equation}
    \mathrm{Tr}\bigl[\Phi_t^{AD}(\rho)\,M\bigr]
    = \mathrm{Tr}\bigl[\rho\,\Phi_t^{*\,AD}(M)\bigr],
\end{equation}
with
\begin{equation*}
    \Phi_t^{*\,AD}(M)
    = \sum_{k=0}^{\infty}F_k^\dagger(t)\,M\,F_k(t),
\end{equation*}
where the Kraus operators \(F_k(t)\) satisfy $\sum_kF_k^\dagger(t)F_k(t)=\mathbb{I}$. This map is completely positive and unital ($\Phi_t^{*\,AD}(\mathbb{I})=\mathbb{I}$), ensuring that any POVM $\{M_j\}$ is assigned to $\{\Phi_t^{*\,AD}(M_j)\}$ while preserving the completeness relation.

For binary photodetection (no click vs. click), consider the ideal POVM  
\begin{equation*}
    M_0 = \lvert0\rangle\langle0\rvert,
    \qquad
    M_1 = \mathbb{I}-\lvert0\rangle\langle0\rvert,
\end{equation*} 
where $M_0$ corresponds to the no-click outcome (vacuum) and $M_1$ to the click event (one or more photons). In terms of detection efficiency, applying the dual map $\Phi_t^{*\,AD}$ to $M_0$ and $M_1$ yields  
\begin{equation}
\begin{aligned}
    M_0(\eta_{\mathrm{tot}}) &= \sum_{n=0}^{\infty}(1-\eta_{\mathrm{tot}})^n\,\lvert n\rangle\langle n\rvert,
    \qquad  \\
    M_1(\eta_{\mathrm{tot}}) &= \mathbb{I}- M_0(\eta_{\mathrm{tot}}),
\end{aligned}
\end{equation}
The factor $(1-\eta_{\mathrm{tot}})^n$ represents the probability of failing to register any of the $n$ incident photons. These operators satisfy  
$M_0(\eta_{\mathrm{tot}})+M_1(\eta_{\mathrm{tot}})=\mathbb{I}$ and thus form a valid POVM for any state $\rho$.

In Bell correlation scenarios, one often restricts to the two-dimensional subspace $\mathrm{span}\{\lvert0\rangle,|\Xi\rangle\}$ \cite{Quintino_2012,araujo2012}. In this case, the effective detection efficiency for the state $\lvert\Xi\rangle$ is
\begin{equation}
    H(\eta_{\mathrm{tot}})
    = \langle\Xi|M_0(\eta_{\mathrm{tot}})|\Xi\rangle
    = \sum_{n=0}^{\infty}(1-\eta_\mathrm{tot})^n\,\bigl|\langle n|\Xi\rangle\bigr|^2,
\end{equation}
This parameter can then be inserted into Bell inequalities to estimate a critical threshold $\eta_c$. However, for any $\eta_{\mathrm{tot}}<\eta_c$, observing a violation is based on the assumption of fair-sampling and does not constitute a loophole-free test.

Therefore, modeling inefficiencies in Bell tests based on amplitude-damping quantum channels and their mathematically equivalent transformations on the states or the measurements underpins experimental strategies. As demonstrated by Araújo et al., incorporating losses directly into measurement operators via the dual channel model allows violations of Bell inequalities even with arbitrarily low photodetection efficiency, using hybrid homodyne–binary detection schemes \cite{araujo2012}. Quintino et al. identified optical states that maximize this violation and established minimum per-mode efficiencies required to sustain it \cite{Quintino_2012}, while Brask and Chaves applied local filters to the states to compensate for inefficiencies modeled by damping channels \cite{Brask_e_Chaves}. Chaves and Brask demonstrated the robustness of single-photon W states in this context \cite{Chaves_Brask_2011}. Teo et al. and Cavalcanti et al. expanded hybrid approaches by combining global-state deformations and measurement operators to maximize violations under realistic conditions \cite{cavalcanti2011,Teo2013}. The choice of modeling losses in the states, the measurements, or both therefore depends on trade-offs between experimental robustness and implementation complexity.

\section{The Instrumental Scenario}
\label{sec:The Instrumental Scenario}

The Bell scenario has long served as the canonical framework for studying nonlocal correlations and testing the limits of classical explanations for quantum phenomena. As we have studied in the previous sections, this scenario assumes strict input independence and spacelike separation. 

However,  many real-world situations — particularly in quantum communication and causal inference — involve asymmetric causal structures and some level of communication. This motivates the study of the instrumental scenario, where the outcome of party $A$ is communicated to party $B$. Originally introduced in classical causal inference \cite{Pearl2009causality}, the instrumental scenario allows one to explore causal constraints beyond locality, including hidden influences and communication. Importantly, its structure enables the derivation of new inequalities and nonclassicality tests that are inequivalent to those of Bell-type setups, and in some cases, can be implemented under more relaxed experimental assumptions. Studying the instrumental scenario thus opens a complementary perspective on quantum correlations and enhances our understanding of how causal structure shapes the boundary between classical and quantum predictions.

To introduce this scenario, let us begin by considering the following situation: given two variables \( A \) and \( B \), one wants to determine whether the correlations between $A$ and $B$ are due to direct causal influences, that is, \( A \) affects directly \( B \) for example, or due to some inaccessible common cause described by a random variable \( \lambda \) \cite{chen2002maximal}. One possible action one can take is to fix a value on the variable \( A \) independently of the value taken by \( \lambda \)  and observe the consequences on the distribution on $B$. This is what is referred to as performing an intervention on variable \( A \).

Unlike passive observations, an intervention involves deliberately modifying the causal setup of an experiment by isolating a specific variable from any external influence and placing it fully under the experimenter’s control. This act breaks any existing correlation between $A$ and $B$ that might be due to a hidden common cause $\lambda$. If, after performing the intervention, the joint distribution satisfies $p(a,b) \neq p(a)p(b)$, it indicates that $A$ exerts a direct causal effect on $B$. Several methods exist to quantify such causal effects but a widely used metric is the so-called Average Causal Effect (ACE) \cite{Instrumental_Miklin} 

\begin{equation}
    ACE_{A\rightarrow B}= \text{max}_{a,a',b} |p(b|\text{do}(a)) -  p(b|\text{do}(a'))|,\label{eq: ACE}
\end{equation}
where the $\text{do}-$probabilities constitute the interventional distribution defined as 

\begin{equation}
p(b|\mathrm{do}(a)) = \sum_{\lambda} p(b|a, \lambda)p(\lambda).
\end{equation}
This definition formally shows that the intervention effectively eliminates any possible external correlation $\lambda$. 

To illustrate the principle of causal intervention, we introduce a pedagogical example based on the São João festival in Northeast Brazil. At these festivals, the presence of a live forró band is common. We wish to determine if this feature causally influences the happiness of the attendees. But then we have the following causal puzzle: \textit{observational data} (data that we observe from past events) reveals a strong correlation between the presence of a forró band (described by a binary variable $F$) and guest happiness (described by another variable $H$). However, a latent variable $\beta$ (in this context also known as confounding factor) may exist: the party's budget. A large budget can both afford a live band (so we have the causal influence $\beta\longrightarrow F$) and improve guest happiness through other means (e.g. higher quality food and drinks; so we have also $\beta\longrightarrow H$). This confounding factor makes it impossible to determine from observation alone whether the forró band has any direct causal effect on happiness $F\longrightarrow H$). The solution is to perform an intervention: to distinguish causation from correlation, one must perform a controlled experiment. By intervening, we would set the value of $F$ independently of the budget. For instance, we could arrange for some parties to have a band and others not to. By measuring the resulting happiness $H$ in both scenarios, we can isolate the true causal effect of the forró band, thereby verifying or refuting a direct causal link $F\longrightarrow H$.

However, performing a direct intervention is not always possible due to ethical, practical, or financial reasons. For instance, one cannot ethically force a group of people to smoke to determine if it causes lung cancer. In such cases where experiments are not feasible, the instrumental scenario provides a powerful alternative using purely observational data. The core idea is to find a naturally occurring variable, the instrument $X$, that satisfies three strict conditions: \textit{i)}  It has a causal effect on the treatment $A$.  \textit{ii)} It only affects the outcome $B$ through the treatment $A$. \textit{iii)}   It is independent of the confounding factor $\lambda$ that influence both $A$ and $B$. 

Returning to our São João party example, a valid instrument $X$ could be the ``unexpected availability of a key musician." This event influences whether the band plays ($F$) but is independent of the party's budget ($\beta$) and does not directly affect guest happiness ($H$). By observing how this ``randomness" in musician availability correlates with guest happiness, we can mathematically isolate the true causal effect of the forró music, approximating the result of a direct experiment without ever having to perform one.  

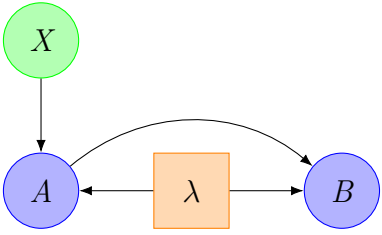
\begin{figure}[h!]
\centering
\begin{tikzpicture}[
    node distance=1.5cm,
    ab circle node/.style={circle, draw=blue, fill=blue!30, minimum size=1cm, inner sep=0pt, font=\large\itshape}, 
    other circle node/.style={circle, draw=green, fill=green!30, minimum size=1cm, inner sep=0pt, font=\large\itshape}, 
    every rectangle node/.style={rectangle, draw=orange, fill=orange!30, minimum size=1cm, inner sep=0pt, font=\large\bfseries}, 
    ->, >={Latex}
]

\node[ab circle node] (A) at (0,0) {A};
\node[ab circle node] (B) [right of=A, xshift=2.5cm] {B};
\node[every rectangle node] (lambda) at ($(A)!0.5!(B)$) {$\lambda$}; 
\node[other circle node] (X) [above of=A, yshift=0.5cm] {X};

\draw[->] (X) -- (A);                 
\draw[->] (lambda) -- (A);           
\draw[->] (lambda) -- (B);           
\draw[->, bend left=40] (A) to (B);  
\end{tikzpicture}
\caption{\justifying \textbf{DAG depicting the causal structure of the instrumental scenario.}}
\label{fig:instrumental}
\end{figure}

The instrumental scenario can also be modeled as a DAG. We present this causal structure  in Fig.~\ref{fig:instrumental}, where any  observational probability distribution compatible with the instrumental scenario causal structure must follow the decomposition 
\begin{equation}
p(a,b|x) = \sum_{\lambda} p(a|x, \lambda)p(b|a, \lambda)p(\lambda).\label{eq: instrumental_cm}
\end{equation}

Notice that Eq.~(\ref{eq: instrumental_cm})  is the classical model equivalent to the HV model Eq.~(\ref{LHV model}) in the Bell scenario. In the same way, Eq.~(\ref{eq: instrumental_cm}) defines a polytope and the scenario can be characterized through the inequalities defining the facets of the polytope. 

If we define  the instrumental scenario by the tuple \( (n_x, n_a, n_b) \) where \( n_x \) are the inputs for Alice, \( n_a  \) the outputs of Alice (inputs of Bob)  and \( n_b \) the  outputs for Bob, then for the scenario $(n_x,2,2)$ with $n_x=2,3,4$ it is known that the polytopes are completely defined by the following inequalities (up to relabeling of the $x$)~\cite{inst_inequalities_x}.
When $X$ is binary, the facets of the polytope are
\begin{equation}
    p(a,b|x) + p(a,b\oplus1|x\oplus1) \leq1,
\end{equation}
where $\oplus$ indicates sum modulo $2$.  These were called Pearl's inequalities in Ref.~\cite{pearl1995causal}. In the case of  a trinary variable $X$ we have the so-called Bonet inequalities~\cite{bonet_inequalities}
\begin{align}
    p(a,b|0) &- p(a,b|1) -p(a\oplus1,1|1)\nonumber\\&-p(a\oplus1,b\oplus1|2) - p(a,b|2) \leq 0,
\end{align}
and for  $n_x=4$ the corresponding inequalities are the Kedagni's inequalities.
\begin{align}
    &p(a, b|0) + p(a\oplus1, b|0) - p(a, b\oplus1|1)\nonumber\\ &- p(a\oplus1, b|1) - p(a, b|2) - p(a\oplus1, b|2)\nonumber\\ &- p(a, b|3) - p(a\oplus1, b\oplus1|3) \leq 0.
\end{align}

Apart from these inequalities characterizing the set of observational correlations in the instrumental scenario, it is interesting to look at the following expression applicable when all variables are binary \cite{GMR_QCI}:
\begin{equation}
    ACE_{A\rightarrow B} \geq
2p(0, 0|0) + p(1, 1|0) + p(0, 1|1) + p(1, 1|1) - 2,\label{eq: ACE_ineq}
\end{equation}
This inequality constitutes a bound to the $ACE$ by considering the correlations of the form Eq.~(\ref{eq: instrumental_cm}). Now the power of the instrumental scenario  is  clear: without performing interventions, only by considering observable distributions we can bound the causal effect between $A$ and $B$. This is the essence of the instrumental scenario. 

In the context of detecting non-classicality, one examines the correlations predicted by quantum theory. This is modeled by replacing the classical latent variable with a shared quantum state, on which each party may perform local measurements corresponding to a POVM. In this scenario, the instrumental (observational) probability is given by 
\begin{equation}
p(a,b|x) = \mathrm{Tr}\left[(M^x_a \otimes N^a_b) \, \rho_{AB}\right].
\end{equation}

While the do-probabilities (interventional) are given by 
\begin{equation}
p(b|\mathrm{do}(a)) = \mathrm{Tr}\left[(I_A \otimes N^a_b) \, \rho_{AB}\right].
\end{equation}

The observational probability distribution resembles that of a Bell scenario, with the crucial difference that Bob's measurement setting is determined by Alice's measurement outcome ($y=a$). In contrast, for an intervention where Bob's measurement is chosen independently of Alice, this causal separation is formally represented by applying the identity map to Alice's system.

Interestingly, in the simplest  \( (2, 2, 2) \) scenario, Pearl's  inequalities remain valid even if the common source is replaced by a quantum state (or even considering a post-quantum box). Unexpectedly, by considering partial entangled states, Ineq.~(\ref{eq: ACE_ineq}) can be violated by quantum mechanics~\cite{PhysRevA.104.L010201}.  Therefore, the following observation is pertinent now: note that when considering only observable distributions (Pearl's inequalities) the classical and quantum sets coincide, meanwhile considering also interventional probabilities (ACE inequality) one is able to witness quantum correlations. Also note that the $ACE$  only considers two values of the whole interventional distribution. Then a natural question that one can ask is whether considering the whole interventional distribution increases the detection power of  quantum correlations and further more, can one build more robust inequalities by considering the observational and the interventional distribution simultaneously. This is the content of Ref.~\cite{ObsInt_Bell}. 

In that article the authors showcase the power of interventions in the aid of detecting non-classicality in the instrumental scenario. The main idea is to consider the observational and interventional distribution together. This defines the vertices of a hybrid polytope as 
\begin{equation}
    \vec{p} =  \vec{p}_{obs} \oplus  \vec{p}_{int}.   
\end{equation}

where the entries $(\vec{p}_{obs})_{a,b,x} = p(ab|x)$ and $(\vec{p}_{int})_{a,b} = p(b|do(a))$ are given by the deterministic strategies of the scenario. The facets of the polytope now will be hybrid inequalities containing do-probabilities and the observational ones. We begin by considering the simplest scenario $(2,2,2)$. For this scenario we have the non-trivial inequality

\begin{equation}\label{EQ15}
I_{222} \geq 0,
\end{equation}

where $I_{222} = p(b|\mathrm{do}(a)) - p(a, b|x) + p(\bar{a}, b|x) + p(a, \bar{b}|x) - p(\bar{a}, \bar{b}|x)$. Here \( x \) are the inputs of Alice, \( a \) are the outputs of Alice and inputs of Bob, and \( b \) are the outputs of Bob. They can take the two values $0$ and $1$. The values with the overline  indicate that they are any value different from the one without the overbar.

Remarkably, quantum correlations can violate this inequality, contrary to the purely observational case. In the next subsection, we review the known results regarding the detection loophole in this scenario and present new results based on alternative non-ideal detector models.

\subsection{Efficiency Requirements in the Instrumental Scenario}\label{map}
The first result regarding inefficiencies in the (2,2,2) instrumental scenario that we present was first obtained in  Ref.~\cite{PhysRevA.104.L010201}. In that work, it was studied the robustness of the ACE inequality~(\ref{eq: ACE_ineq}). Two imperfect detector models were considered. The first is the \textit{trusted detector} model, which assumes each detector has a fixed, state-independent efficiency $\eta$. This is a physically well-motivated assumption, for instance, in the case of photon detectors measuring polarization-encoded qubits. The second, more general model is that of \textit{untrusted detectors}, where the detection efficiency may depend on the specific quantum state being measured. In this model the constraint that the overall probability of a successful joint detection must be constant for any measurement setting $x$ is imposed. This is enforced by the condition
\begin{equation}
    \sum_{a,b} p(a, b|x) = \eta^2, \quad \forall x.
\end{equation}
This ensures that any malicious behavior that alters the total detection rate can be readily identified. 
In the analysis, it was concluded that the minimum efficiency for having a quantum violation free of the detection loophole is $\eta \approx 0.96$. It was also defined the inequality 
\begin{align}
    ACE_{A\rightarrow B} \geq
2p(0, 0|0) &+ p(1, 1|0) + p(0, 1|1) + p(1, 1|1)\nonumber \\&- 1-\eta^2,\label{eq: ACE_ineq_cao}
\end{align}
It was proven  that this inequality allows to lower the efficiency to $.92$. As we already saw in Sec.~\ref{sec:The Bell scenario}, these efficiencies are much higher than those for the corresponding Bell scenario ($\eta \approx 2/3)$ \cite{Eberhard93}.

We now consider imperfect detectors in the instrumental scenario and reproduce the minimum detection efficiency thresholds required for a loophole-free quantum violation of inequality (\ref{EQ15})~\cite{ObsInt_Bell}. To this end, one can define a map valid for any nonsignaling Bell distribution \(p_{\rm Bell}(a,b| x,y)\), and which directly yields the instrumental probabilities:

\begin{align}\label{eq18}
&p(a,b|x) = p_{\rm Bell}\bigl(a,b|x,\,y=a\bigr), \quad \forall a,b,x \\ \nonumber
&p\bigl(b| \mathrm{do}(a)\bigr) = \sum_{a'} p_{\rm Bell}\bigl(a',b| x,\,y=a\bigr), \quad \forall a,b,x
\end{align}
as discussed in \cite{GMR_QCI}. This mapping preserves all local, quantum, and post-quantum features, since the nonsignaling condition in \(y\) ensures that the choice \(y = a\) yields marginal distributions identical to those for any other \(y\), thus guaranteeing consistency and independence of the instrumental marginals. 

Next, we introduce the detection efficiencies \(\eta_{1}\) (Alice) and \(\eta_{2}\) (Bob), each varying in \([0,1]\). In the independent detection model, each measurement attempt results in a “click” with probability \(\eta_i\) and a “no-click” with probability \(1 - \eta_i\). 

\subsubsection{Modeling Non-Detection Events by Absorption into Another Output}
\label{sec: instrumental_Model1}

In realistic Bell-type experiments, imperfect detectors often fail to register an outcome, leading to no-detection events that must be incorporated into the theoretical model. As mentioned in Sec.~\ref{sec:ND_models}, one approach to address this issue is to absorb these events into effective outcomes, thus redefining the observed statistics. This section formalizes such a model by expressing the modified probabilities in terms of ideal Bell distributions and detection efficiencies. 

 Let us consider Eq.~(\ref{eq:abs_1}), Eq.~(\ref{eq:abs_2}) and the analogous equation for $A$ together with the normalization constraint. We can map these equations into the instrumental scenario using the map given by Eqs.~(\ref{eq18}), obtaining \cite{ObsInt_Bell}:

\begin{align}\label{eq20}
p_{\eta_1 \eta_2}(a,b|x) &= \eta_1 \eta_2 \, p(a,b|x), \\ \nonumber
p_{\eta_1 \eta_2}(a,\hat{b}|x) &= \eta_1 \eta_2 \, p(a,\hat{b}|x) \\ \nonumber & + \eta_1(1 - \eta_2) \, p(a|x), \\ \nonumber
p_{\eta_1 \eta_2}(\hat{a},b|x) &= \eta_1 \eta_2 \, p(\hat{a},b|x) \\ \nonumber & + (1 - \eta_1)\eta_2 \, p(b|\mathrm{do}(\hat{a})), \\ \nonumber
p_{\eta_1 \eta_2}(\hat{a},\hat{b}|x) &= \eta_1 \eta_2 \, p(\hat{a},\hat{b}|x) + \eta_1(1 - \eta_2) \, p(\hat{a}|x) \notag \\ \nonumber
\quad &+ (1 - \eta_1)\eta_2 \, p(\hat{b} | \mathrm{do}(\hat{a})) + (1 - \eta_1)(1 - \eta_2), \nonumber
\end{align}
The corresponding expressions for the interventional do-probabilities are:

\begin{align}\label{eq21}
p_{\eta_2}(b | \mathrm{do}(a)) &= \eta_2 \, p(b | \mathrm{do}(a)), \\ \nonumber
p_{\eta_2}(\hat{b} | \mathrm{do}(a)) &= \eta_2 \, p(\hat{b} | \mathrm{do}(a)) + (1 - \eta_2),
\end{align}
The instrumental scenario can be interpreted as a postselection case of the Bell scenario where the input \(y\) is forced to match Alice’s output \(a\), therefore it is expected that the efficiency thresholds required for violating Bell inequalities in the instrumental setting are not lower than in the original Bell setting.

\begin{figure}[!h]
            \centering           
            \includegraphics[scale=0.57]{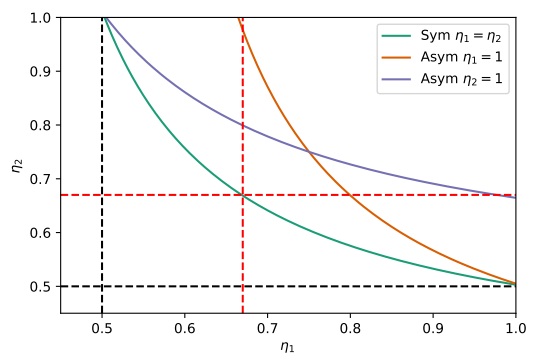}
            \caption{ \justifying \textbf{Violation of the $I_{222}$ absorbing the non-detection events in \(\hat{a}=1\) and \(\hat{b}=0\).} On the other hand, in the figure, we show the function \(\eta_2(\eta_1)\) that identifies the region where the inequality \(I_{222}\) cannot be violated, for the case where \(\hat{a} = 1\) and \(\hat{b} = 0\). In the symmetric configuration (green curve), the minimum required efficiency is \(\eta_1 = \eta_2 = 0.67\) and the curves were obtained with the optimal measurements and states reaching this efficiency. In the asymmetric case with a perfect detector (\(\eta_1 = 1\)), the minimum bound for the other detector is \(\eta_2 = 0.5\); conversely, for \(\eta_2 = 1\), the same minimum \(\eta_1 = 0.5\) is obtained. These efficiency bounds are indicated by black dashed lines. The orange and purple curves were obtained using the settings and states for the optimal efficiencies \(\eta_2 = 0.51\) and \(\eta_1 = 0.51\), respectively.}
            \label{fig4}
    \end{figure}

\begin{figure}[!h]
            \centering           
            \includegraphics[scale=0.57]{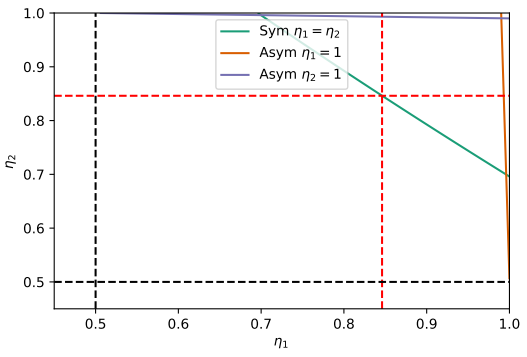}
            \caption{\justifying \textbf{Violation of the $I_{222}$ absorbing the non-detection events in \(\hat{a}=1\) and \(\hat{b}=1\).} In the Figure, we present the curve \(\eta_2(\eta_1)\) indicating the region where no violation of the inequality \(I_{222}\) is possible. In the symmetric case (depicted by the green curve), the minimum efficiency required for a violation is found to be \(\eta_1 = \eta_2 = 0.84\). In the asymmetric scenario where one detector is perfect (i.e., \(\eta_1 = 1\)), the corresponding threshold for the other detector is \(\eta_2 = 0.5\); the same value \(\eta_1 = 0.5\) is required if \(\eta_2 = 1\). These critical efficiency points are represented by black dashed lines. The orange and purple curves were obtained using the measurement and states for the optimal efficiencies \(\eta_2 = 0.51\) and \(\eta_1 = 0.51\), respectively. The green curve was obtained with the optimal measurements and state that give the minimum efficiency of $0.84$.}
            \label{fig3}
    \end{figure}

Interestingly,  when we consider the symmetry of inequality (\ref{EQ15})

\begin{equation*}
    I_{222} =  p(1|do(0))-p(0, 1|1)+p(0, 0|0)+p(1, 1|0)-p(1, 1|1),
\end{equation*}
if tested using the noisy probabilities from Eq.~(\ref{eq20}) and Eq.~(\ref{eq21}), one obtains results fully consistent with the Bell scenario. This is explained by noting that through the map given in Eq.~(\ref{eq18}), the inequality $I_{222}$ is equivalent to a CHSH inequality \cite{ObsInt_Bell}. Note that given a fixed representative of the inequalities (\ref{EQ15}), the efficiency depends on which output we absorb as in the Bell case.  In particular, when non-detection events are absorbed in the outcomes \(\hat{a} = 1\) and \(\hat{b} = 0\), by a numerical optimization over all measurement and states the minimal efficiency in the symmetric case is found to be \(\eta_1 = \eta_2 = 0.6756\), as shown in Fig.~\ref{fig4}. On the other hand if we absorb  \(\hat{a} = 1\) and \(\hat{b} = 1\) for example, the minimum efficiency in the symmetric case turns to be \(\eta_1 = \eta_2 = 0.84\), this is shown in  Fig. \ref{fig3}.

We also considered the asymmetric case, where one detector is perfect and the other is not. In Fig.~\ref{fig4} and Fig.~\ref{fig3}  we show the curves, orange and purple, relating $\eta_1$ and $\eta_2$  when we fix the measurements and states of the optimal strategy for $\eta_1=1$ and $\eta_2=1$ respectively. We emphasize that these curves are identical to those in Figs.~\ref{fig2} and~\ref{fig1} due to the formal mapping between the scenarios discussed in Sec. \ref{map}", and the result was first obtained in Ref.~\cite{ObsInt_Bell}.

\subsubsection{Modeling Non-Detection Events as an Additional Output}

Here we complement the previous known results with a previously un-annalyzed additional output model. In this second model, non-detection events are treated as explicit additional outcomes, denoted by $\varnothing$, which represent the absence of a detection click. In photonic experiments, such a modeling choice can be realized by assuming that both Alice and Bob are equipped with two photodetectors sensitive to orthogonal polarizations. The presence of these additional outcomes extends the non-ideal Bell scenario $(n_x = 2, n_a = 2;\, n_y = 2, n_b = 2)$ to the enlarged configuration $(n_x = 2, n_a = 3;\, n_y = 2, n_b = 3)$, where $n_x$ and $n_y$ denote the number of input settings, and $n_a$, $n_b$ the number of output values available to Alice and Bob, respectively.

From Sec.~\ref{sec:ND_models}, the observed probabilities under detection inefficiencies relate to the ideal Bell probabilities as follows:

\begin{align}
&p^{\text{Bell}}_{\eta_1 \eta_2}(a, b | x, y) = \eta_1 \eta_2\, p^{\text{Bell}}(a, b | x, y), \nonumber \\
&p^{\text{Bell}}_{\eta_1 \eta_2}(a, \varnothing | x, y) = \eta_1(1 - \eta_2)\, p^{\text{Bell}}(a | x), \nonumber \\
&p^{\text{Bell}}_{\eta_1 \eta_2}(\varnothing, b | x, y) = (1 - \eta_1)\eta_2\, p^{\text{Bell}}(b | y), \nonumber \\
&p^{\text{Bell}}_{\eta_1 \eta_2}(\varnothing, \varnothing | x, y) = (1 - \eta_1)(1 - \eta_2), \label{eq:bell_model2}
\end{align}
In this context, the variables $a$ and $b$ denote only the conclusive results, excluding non-detection events $\varnothing$. The marginal probability $p^{\text{Bell}}(a | x)$ is thus defined as the sum over all valid outcomes of Bob: $p^{\text{Bell}}(a | x) = \sum_{b \neq \varnothing} p^{\text{Bell}}(a, b | x, y)$, which remains well-defined since $\varnothing$ does not belong to the ideal outcome set. When mapping to the instrumental scenario, in which Alice's output is used as Bob's input, conceptual complications arise. Specifically, expressions such as $p(b | \text{do}(\varnothing))$ are not well defined in ideal theory, since $\varnothing$ is not a valid outcome. 

The resulting expressions under the instrumental mapping become

\begin{align}
&p_{\eta_1 \eta_2}(a, b | x) = \eta_1 \eta_2\, p(a, b | x), \nonumber \\
&p_{\eta_1 \eta_2}(a, \varnothing | x) = \eta_1(1 - \eta_2)\, p(a | x), \nonumber \\
&p_{\eta_1 \eta_2}(\varnothing, b | x) = (1 - \eta_1)\eta_2\, p(b | \text{do}(\varnothing)), \nonumber \\
&p_{\eta_1 \eta_2}(\varnothing, \varnothing | x) = (1 - \eta_1)(1 - \eta_2) \sum_b p(b | \text{do}(\varnothing)), \label{eq:instrumental_model2}
\end{align}
The last two expressions in Eqs.~\eqref{eq:instrumental_model2} are ill-defined due to their dependence on interventions conditioned on the non-detection event $\varnothing$, which lies outside the support of the ideal scenario.

To resolve the ill-defined nature of conditional probabilities involving non-detection events (e.g., $p(b | \text{do}(\varnothing))$), we consider two physically motivated modeling strategies that preserve internal consistency within the instrumental scenario.

\paragraph{Perfect Detection Assumption on Alice's Side.}
Assuming that Alice's detectors operate with unit efficiency, that is, $\eta_1 = 1$, all her outcomes are conclusively registered and the event $\varnothing$ never occurs on her side. Consequently, the observed joint and interventional probabilities reduce to:
\begin{align}
&p_{1 \eta_2}(a, b | x) = \eta_2\, p(a, b | x), \nonumber \\
&p_{1 \eta_2}(a, \varnothing | x) = (1 - \eta_2)\, p(a | x), \nonumber \\
&p_{1 \eta_2}(\varnothing, b | x) = 0, \nonumber \\
&p_{1 \eta_2}(\varnothing, \varnothing | x) = 0,
\end{align}

with the corresponding interventional (do-) probabilities given by:
\begin{align}
&p_{1 \eta_2}(b | \text{do}(a)) = \eta_2\, p(b | \text{do}(a)), \nonumber \\
&p_{1 \eta_2}(\varnothing | \text{do}(a)) = 1 - \eta_2.
\end{align}

\paragraph{ Hybrid Model with Partial Absorption of Non-Detection.}
Alternatively, one may consider a hybrid approach in which non-detection events on Alice’s side are absorbed into a valid outcome, while Bob's side includes $\varnothing$ as an explicit additional outcome. The observed statistics in this case obey 
\begin{align}
&p_{\eta_1 \eta_2}(a, b | x) = \eta_1 \eta_2\, p(a, b | x), \nonumber \\
&p_{\eta_1 \eta_2}(a, \varnothing | x) = \eta_1(1 - \eta_2)\, p(a | x), \nonumber \\
&p_{\eta_1 \eta_2}(\hat a, b | x) = \eta_1 \eta_2\, p(\hat a, b | x) + (1 - \eta_1)\eta_2\, p(b | \text{do}(\hat a)), \nonumber \\
&p_{\eta_1 \eta_2}(\hat a, \varnothing | x) = 1 - p_{\eta_1 \eta_2}(a, b | x) -p_{\eta_1 \eta_2}(a, \varnothing | x) \nonumber \\&- p_{\eta_1 \eta_2}(\hat a, b | x).
\end{align}

The corresponding do-probabilities take the form:

\begin{align}
&p_{\eta_2}(b | \text{do}(a)) = \eta_2\, p(b | \text{do}(a)), \nonumber \\
&p_{\eta_2}(\varnothing | \text{do}(a)) = 1 - \eta_2.
\end{align}

By explicitly incorporating $\varnothing$ as a well-defined outcome (or absorbing it appropriately), one can preserve the operational meaning of interventional statistics and ensure compatibility with causal and nonlocality-based inequalities.

\begin{figure}[!h]
            \centering           
            \includegraphics[scale=0.4]{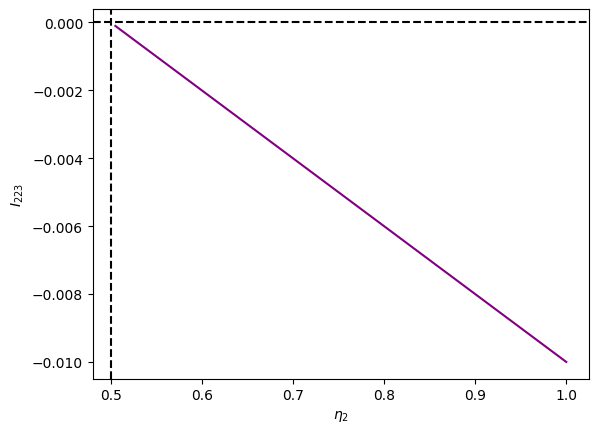}
            \caption{ \justifying \textbf{${I}_{223}$ for model \textit{a.} with \(\eta_1 = 1\)}. In this case, an ideal detector is assumed on side A, while a third outcome \(\varnothing = 2\) is included to account for inefficiencies on side B. The plotted curve corresponds to an optimization of \(I_{223}\) under the constraint \(\eta = 0.51\). The intersection of the dashed lines indicates that for \(\eta_2 = 0.5\), the inequality reaches its classical bound of zero, thus no longer detecting quantum correlations. The curve was obtained with the optimal measurements and state that give the minimum critical efficiency.}
            \label{fig5}
    \end{figure}

In both approaches, Bob is assumed to have access to three possible outcomes, while Alice maintains only two. The instrumental inequality under investigation in this context is:
\begin{equation}
\begin{aligned}
I_{223} =\ & p(00 | 1) - p(01 | 0) + p(02 | 1) \\
           & - p(11 | 0) + p(11 | 1) + p(1 | \text{do}(0)) \geq 0,
\end{aligned}
\end{equation}
This inequality corresponds to a facet of the classical hybrid polytope of the instrumental scenario $(2,2,3)$. In the first case, where Alice’s detectors are assumed to be ideal, we optimize over the detection efficiency $\eta_2$ considering all dichotomic measurements and qubit states, seeking the minimum value that still permits a quantum violation of $I_{223}$. Fig.~\ref{fig5} illustrates the behavior of $I_{223}$ as a function of $\eta_2$, for an optimized choice of measurements and quantum states with $\eta_2 = 0.51$.

In the second hybrid modeling approach, we recover the same thresholds previously observed in Figs. (\ref{fig4}) and (\ref{fig3})  for the absorption model (see Sec.~\ref{sec: instrumental_Model1}). Specifically, the minimum efficiency required for a violation is $\eta = 0.67$ in the symmetric case ($\eta_1 = \eta_2$), and $\eta = 0.5$ in the fully asymmetric configuration, where one detector is ideal and the other not. 

            \label{fig7}

In addition to the two solutions presented above, a more general one can be obtained by intoducing an additional outcome  on Alice’s side, and both an extra input and output are allowed on Bob’s side within the ideal configuration. The extra outputs will correspond to the no-click event.  This leads us to consider inequalities of the form:

\begin{equation}
I_{233} \geq 0.
\end{equation}

To explore this regime, we analyzed the complete polytope associated with the instrumental scenario $(2,3,3)$, which contains more than 900 nontrivial facets. Among these, only 64 facets remain unviolated by quantum statistics. In particular, although Alice still selects between two measurement settings, the variable $a$ now ranges over three possible values, reflecting the expanded outcome space.

In the symmetric detection efficiency regime ($\eta_1 = \eta_2$), the lowest efficiency for which a quantum violation was detected is approximately $\eta_1 = \eta_2 \approx 0.9052$. This bound was achieved using, for instance, the following inequality:
\begin{equation}
\begin{aligned}\label{eq31}
I_{233} =\ & -p(00 | 1) - p(01 | 0) - p(02 | 1) + p(0 | \text{do}(1)) \\
           & - p(10 | 0) - p(10 | 1) + 1.
\end{aligned}
\end{equation}

\begin{figure}[!h]
            \centering           
            \includegraphics[scale=0.55]{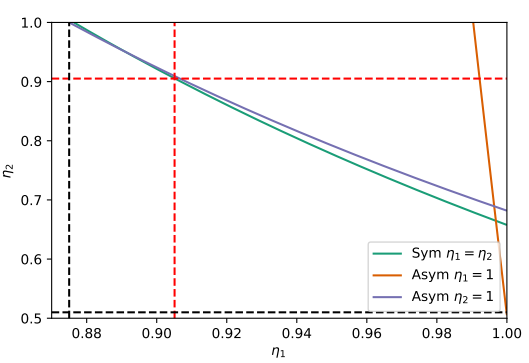}
            \caption{ \justifying  \textbf{Efficiencies in the $I_{233}$ inequality.} The non-detection event in $A$ and $B$ corresponds to the $a=2$ and $b=2$  outputs respectively. The figure displays the values of \( \eta_2(\eta_1) \) for which no violation of the inequality \( I_{233} \) occurs. These correspond to the region below the curves. In the symmetric case (green curve), the minimum efficiency required is \( \eta_1 = \eta_2 = 0.90 \). In the asymmetric configuration with a perfect detector on side $A$ (\( \eta_1 = 1 \)), the minimum value for \( \eta_2 \) is $0.51$. Conversely, if \( \eta_2 = 1 \) is assumed, the threshold for \( \eta_1 \) becomes $0.87$. These critical values are indicated by the black dashed lines. The purple and red curves correspond to the optimal strategy obtained for \( \eta_1 = 0.87 \) and \( \eta_2 = 0.51 \), respectively.}
            \label{fig9}
    \end{figure}

In the asymmetric case, the minimal detection efficiency required depends on which party is assumed to have a perfect detector. When Alice's efficiency is fixed at $\eta_1 = 1$, the minimum efficiency required for Bob is $\eta_2 \approx 0.51$. In contrast, assuming $\eta_2 = 1$, the minimum efficiency achievable on Alice’s side is limited to $\eta_1 \approx 0.8750$. These results are depicted in Fig.~\ref{fig9}, based on the inequality (\ref{eq31}).

\section{The prepare-and-measure Scenario}
\label{sec:The Prepare-and-Measure Scenario}

In the previous sections, we introduced what we refer to as device-independent scenarios. However, in practical quantum experiments, it is often necessary to assume certain physical properties of the setup. In this section, we describe the prepare-and-measure (PAM) scenario, also known as the semi-device-independent scenario due to the assumption of a bound on the system's dimension.

Discussions about such dimensional restrictions began in the context of Bell scenarios, where Bell-type correlations were first used to establish lower bounds on the Hilbert space dimension required to reproduce observed statistics~\cite{brunner2008}. A more direct formulation of a scenario with an explicit dimensional constraint was later proposed by Gallego \textit{et al.}, who introduced the framework now known as the PAM scenario~\cite{gallego2010}. This framework was further developed by Navascués and Vértesi, who provided semidefinite programming (SDP) techniques to bound the set of finite-dimensional quantum correlations~\cite{navascues2015}. The PAM scenario also offers a natural setting to explore how system dimensionality impacts information processing, particularly in the security of semi-device-independent quantum key distribution and randomness expansion protocols~\cite{pawlowski2011, li2011, li2012}.

Experimental efforts have demonstrated violations of dimension witness inequalities in prepare-and-measure scenarios. For example, photon pairs entangled in polarization and orbital angular momentum were used to certify system dimensionality~\cite{hendrych2012}. Semi-device-independent tests have distinguished classical from quantum systems based on measurement statistics~\cite{ahrens2012} and extended these methods to cryptographic applications~\cite{ahrens2014}. 

In a typical PAM scenario, two devices are involved: a preparation device and a measurement device. Upon receiving an input \(X=x\), drawn from a finite set of choices, the preparation device emits a physical system whose state depends on \(x\). This system is then transmitted to the measurement device, which, upon receiving a separate input \(Y=y\), performs a measurement and produces an output \(B=b\). The statistics of the observed outcomes \(p(B=b|X=x,Y=y) = p(b|x,y)\) thus depend on the choice of inputs and the nature of the systems exchanged between the devices. The underlying assumption in the semi-device-independent setting is that the physical system used to communicate between the devices is of bounded dimension, which constrains the set of possible correlations and enables certification tasks under minimal assumptions.

\begin{figure}[h!]
\centering
\begin{tikzpicture}[
    node distance=1.5cm,
    ab circle node/.style={circle, draw=blue, fill=blue!30, minimum size=1cm, inner sep=0pt, font=\large\itshape}, 
    other circle node/.style={circle, draw=green, fill=green!30, minimum size=1cm, inner sep=0pt, font=\large\itshape}, 
    every rectangle node/.style={rectangle, draw=orange, fill=orange!30, minimum size=1cm, inner sep=0pt, font=\large\bfseries}, 
    ->, >={Latex}
]

\node[ab circle node] (M) at (0,0) {M};
\node[ab circle node] (B) [right of=M, xshift=2.5cm] {B};
\node[every rectangle node] (Lambda) at ($(M)!0.5!(B)$) {$\lambda$}; 
\node[other circle node] (X) [above of=M, yshift=0.5cm] {X};
\node[other circle node] (Y) [above of=B, yshift=0.5cm] {Y};

\draw[->] (X) -- (M);                 
\draw[->] (Y) -- (B);                 
\draw[->] (Lambda) -- (M);           
\draw[->] (Lambda) -- (B);           
\draw[->, bend left=40] (M) to (B);  
\end{tikzpicture}
\caption{\justifying \textbf{DAG depicting the causal structure of a prepare-and-measure scenario with correlated preparation and measurement devices.}}
\label{fig:PAM}
\end{figure}
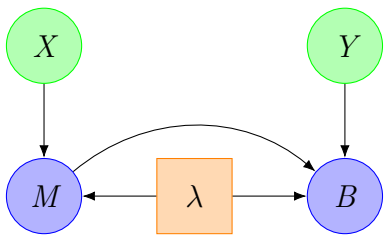

As in previous sections, the classical model for the PAM scenario can be formalized using a causal perspective based on a directed acyclic graph (DAG)~\cite{davide2020}. In this representation (see Fig. \ref{fig:PAM}), the preparation input variable \(X\) influences a hidden message \(M\), which is sent to the measurement device. The measurement input variable \(Y\) and a shared hidden variable \(\lambda\) jointly determine the output \(B\), as well as possibly influencing the preparation. This corresponds to a model with correlated devices, described by the decomposition
\begin{equation}
p(b|x,y) = \sum_{\lambda, m} p(b|m, y, \lambda) \, p(m|x, \lambda) \, p(\lambda).
\label{eq: pam_classical}
\end{equation}

However, an alternative causal structure can describe a classical PAM scenario when different assumptions are made about the devices. If the preparation and measurement devices are assumed to be independent, the hidden variable \(\lambda\) is split into two uncorrelated parts \(\mu\) and \(\nu\), respectively associated with each device. The causal model then satisfies the decomposition
\begin{equation}
    p(b|x,y) = \sum_{\mu, \nu, m} p(b|m, y, \nu)\, p(m|x, \mu)\, p(\mu)\, p(\nu),
\end{equation}
which implies a nonlinear independence constraint \(p(\mu, \nu) = p(\mu) p(\nu)\). This leads to a nonconvex set of correlations, making the analysis of such models more subtle~\cite{bowles2014, davide2020}. To analyze detection efficiencies in this work, we shall focus on the correlated-device model.

We can also consider a quantum version of the PAM scenario, in which the preparation device prepares quantum states \(\rho_x\) depending on the input \(x\). These states are transmitted to the measurement device, which performs a measurement determined by the input \(y\). The measurement is described by a set of positive operator-valued measure (POVM) elements \(\{M_{b|y}\}\), where \(b\) labels the possible outcomes. According to Born's rule,
\begin{equation}
    p(b|x,y) = \mathrm{Tr}\left( \rho_x M_{b|y} \right).
\end{equation}

When the preparation and measurement devices are allowed to share correlations, alternative quantum descriptions arise. For example, the devices may share a quantum state, enabling quantum correlations, while the preparation process itself remains classical~\cite{pawlowski2010}. A further extension includes scenarios where both the shared resource and the communicated systems are quantum, as in the superdense coding protocol~\cite{bennett92}. Although such configurations are of theoretical interest, we will not consider them in this work.

\subsection{Inequalities for Dimension Witnessing in PAM Scenarios}
We now turn our attention to dimension witnesses, which are inequalities associated with the classical description of the PAM scenario with correlated devices, as given in Eq.~\eqref{eq: pam_classical}. While we do not provide a detailed derivation of these inequalities here, the method for obtaining them by listing all extremal points of the polytope defined by the causal structure in Fig.~\ref{fig:PAM} can be found in the appendix of Ref.~\cite{davide2020}.

To illustrate the concept of dimension witnesses, we consider the most elementary prepare-and-measure configuration that features a nontrivial bound on the dimension of the communicated system. We consider the case where the communicated system is classical and of dimension 2, so that the message \(m\) takes binary values. The preparation input is taken from three settings, \(x \in \{0,1,2\}\), the measurement device performs one of two possible measurements, \(y \in \{0,1\}\), and the output is binary, \(b \in \{0,1\}\). In this setting, the only nontrivial dimension witness inequality is the one introduced in Ref.~\cite{gallego2010}, given by
\begin{equation}
    S_3 = E_{00} + E_{01} + E_{10} - E_{11} - E_{20} \leq 3,
    \label{eq: pam_s3}
\end{equation}
where the expectation values are defined as \(E_{xy} = p(0|x,y) - p(1|x,y)\). This inequality must be satisfied by any classical model in which the communicated message has dimension at most 2. However, quantum systems of the same dimension (qubits) can violate this bound, reaching a maximal value of \(S_3 = 1 + 2\sqrt{2} \approx 3.828\). The violation of such an inequality is precisely what justifies calling it a dimension witness, as it certifies, in this case, that the system being communicated cannot be a classical bit. In more general scenarios, different input and output configurations, along with varying bounds on the message dimension, can give rise to other types of dimension witness inequalities.

Another common scenario with message dimension 2 is defined by \(2^n\) preparation settings, where the input \(x\) corresponds to an \(n\)-bit string, i.e., \(x \in \{0,1\}^n\). For each preparation \(x\), there are \(n\) possible measurement choices, labeled by \(y \in \{0, \ldots, n-1\}\), and the measurement outcomes are binary, \(b \in \{0,1\}\). This scenario is closely related to \(n \rightarrow 1\) quantum random access codes (QRACs), where an \(n\)-bit string is encoded into a single-qubit system to enable retrieval of any one of the input bits with high probability~\cite{pawlowski2011, li2011, li2012, Ambainis2008}. A typical dimension witness for this scenario can be written as
\begin{equation}
    T_n = \sum_{x \in \{0,1\}^n} \sum_{y=0}^{n-1} (-1)^{x_y}\,p(0|x,y),
    \label{eq: pam_Tn}
\end{equation}
where \(x_y\) denotes the \(y\)-th bit of the input string \(x\). For \(n=2\), the classical and quantum bounds for the dimension witness \(T_n\) are known exactly: 
\(T_2^C = 2\) and \(T_2^Q = 2\sqrt{2} \approx 2.8284\). 
For \(n > 2\), the optimal values have been established through numerical optimization. Specifically, for \(n=3\), we have \(T_3^C = 6\) and \(T_3^Q \approx 6.9282\); for \(n=4\), \(T_4^C = 12\) and \(T_4^Q \approx 15.4548\); and for \(n=5\), \(T_5^C = 30\) and \(T_5^Q \approx 34.1725\). These results are reported in Refs.~\cite{Ambainis2008, li2012}.

Finally, to describe a general dimension witness in a PAM scenario, we consider finite sets of preparations, measurements, and outputs denoted by \(\mathcal{X}\), \(\mathcal{Y}\), and \(\mathcal{B}\). The behavior of the system is characterized by the set of conditional probabilities \(\mathbf{p} = \{p(b|x,y)\}_{b \in \mathcal{B}, x \in \mathcal{X}, y \in \mathcal{Y}}\). A dimension witness is defined as a linear functional acting on this behavior:
\begin{equation}
    W(\mathbf{p}) = \sum_{x \in \mathcal{X}} \sum_{y \in \mathcal{Y}} \sum_{b \in \mathcal{B}} c_{b|x,y}\, p(b|x,y),
    \label{eq:PAM_W}
\end{equation}
where the coefficients \( c_{b|x,y} \in \mathbb{R} \) define the specific witness. For a fixed dimension \( d \), there exists a classical bound \( W_d^{\mathrm{C}} \), which is the maximum value that \( W(\mathbf{p}) \) can attain when the communicated system is classical and has dimension at most \( d \). If the observed behavior satisfies \( W(\mathbf{p}) > W_d^{\mathrm{C}} \), one can conclude that the system's dimension exceeds \( d \), or that it must be quantum. This general formulation encompasses all previous examples.

\subsection{Efficiency Requirements in the PAM Scenario}
Following the structure of the previous sections that addressed detection inefficiencies in Bell and instrumental scenarios, this section focuses on the PAM scenario. Modeling detection inefficiencies in PAM protocols helps represent experimental imperfections and accurately assess dimension witness tests. Detection losses affect the observed statistics and, if not properly accounted for, can lead to false conclusions about the system's dimension. Additionally, from a security perspective, such inefficiencies can be exploited to simulate dimension violations and compromise protocol integrity. While these security aspects are important, they are beyond the scope of this work; for an in-depth analysis of detection loophole attacks and mitigation strategies in semi-device-independent protocols, see Refs.~\cite{DallArno2015, Mironowicz2021}.

This section explores different approaches for modeling detection inefficiencies and examines their impact on the violation of dimension witness inequalities. We consider three main strategies: one where non-detection events are interpreted as absorptions that reduce the total probability of obtaining valid outcomes; another where such events are explicitly represented as an additional output of the measurement device; and a third approach that accounts for imperfections in the state preparation. Our treatment follows the methods introduced in Refs.~\cite{li2015, dallarno2012}, where the robustness of dimension witnesses is analyzed under similar assumptions.

\subsubsection{Modeling Non-Detection Events by Absorption into Another Output}
\label{sec: pam_absorption}
Like in device-independent scenarios (see Secs.~\ref{sec:ND_models} and \ref{sec: instrumental_Model1} for the analogous approach in Bell and instrumental scenarios, respectively), a common method to model non-detection events is to absorb them into one of the existing output labels, effectively redefining the observed statistics without increasing the cardinality of the output set. The following describes how this can be done in the PAM scenario.

Let us now suppose that the detector associated with measurement input \( y \) has an efficiency \( \eta_y \in [0,1] \). This means that a detection event occurs with probability \( \eta_y \), while with probability \( 1 - \eta_y \) no outcome is recorded. Under the absorption model, we assume that whenever no detection occurs, the device outputs a fixed outcome \( \hat b \in \mathcal{B} \). This assumption leads to the following modified observed conditional probabilities:
\begin{equation}
\tilde{p}(b|x,y) =
\begin{cases}
\eta_y\, p(b|x,y), & \text{if } b \neq \hat b, \\
\eta_y\, p(\hat b|x,y) + (1 - \eta_y), & \text{if } b = \hat b,
\end{cases}
\end{equation}
where \( p(b|x,y) \) are the ideal, lossless probabilities. In other words, the probabilities of outcomes \( b \neq \hat b \) are scaled down by the detection efficiency \( \eta_y \), while the missing weight \( 1 - \eta_y \) is added to the fixed outcome \( \hat b \). This can be equivalently expressed in a more compact form as
\begin{equation}
\tilde{p}(b|x,y) = \eta_y\, p(b|x,y) + (1 - \eta_y)\, \delta_{b,\hat b}.
\end{equation}

In this context, the main objective is to assess whether the observed conditional probability distribution \(\mathbf{\tilde{p}} = \{\tilde{p}(b|x,y)\}_{b,x,y}\), affected by detection inefficiencies, remains compatible with classical or quantum correlations under dimension constraints. According to the absorption model, the corresponding dimension witness value is computed as \( W(\mathbf{\tilde{p}}) \), obtained by evaluating the linear functional defined in Eq.~\eqref{eq:PAM_W} on \(\mathbf{\tilde{p}}\):
\begin{equation}
    W(\mathbf{\tilde{p}}) = \sum_{b,x,y}  c_{b|x,y}\, \tilde{p}(b|x,y).
    \label{eq:pam_tildeW}
\end{equation}
Substituting the explicit form of \(\tilde{p}(b|x,y)\) into the expression yields
\begin{equation}
    W(\mathbf{\tilde{p}}) = \sum_{b,x,y} \eta_y\, c_{b|x,y}\, p(b|x,y) + \sum_{x,y} (1 - \eta_y)\, c_{\hat b|x,y}.
\end{equation}
In other words, the observed witness value under detection inefficiencies corresponds to a weighted sum of the ideal witness evaluated on the lossless behavior, rescaled by the detection efficiencies \(\eta_y\), plus an additive bias term arising from the assignment of non-detection events to the fixed output \(\hat b\). This additive term depends solely on the coefficients \(c_{\hat{b}|x,y}\) of the witness associated with the chosen outcome \(\hat b\). Consequently, the presence of losses effectively modifies both the slope and the offset of the dimension witness value, potentially reducing its violation or, in some cases, even artificially increasing it depending on the choice of \(\hat b\) and the structure of the coefficients \(c_{b|x,y}\). An important special case occurs when the assigned outcome $\hat b$ does not appear in the inequality, i.e., $c_{\hat b|x,y} = 0$ for all $x,y$. In this case, the bias term vanishes and the witness value reduces to
\begin{equation}
W(\mathbf{\tilde{p}}) = \sum_{y} \eta_y\, W_y(\mathbf{p}),
\end{equation}
where $W_y(\mathbf{p}) = \sum_{b,x} c_{b|x,y}\, p(b|x,y)$ denotes the contribution from setting $y$ to the \emph{lossless} witness value $W(\mathbf{p}) = \sum_{y} W_y(\mathbf{p})$. If $\eta_y = \eta$ is constant across all settings, this further simplifies to $W(\mathbf{\tilde{p}}) = \eta\, W(\mathbf{p})$, so that losses merely rescale the lossless witness value.

Based on Ref.~\cite{li2015}, let us now consider a concrete example of the dimension witness introduced in Eq.~\eqref{eq: pam_Tn} for the case where $ n=2$. In this scenario, the preparation inputs correspond to all bit strings $ x \in \{00, 01, 10, 11\}$ , while both the outcomes and measurement settings are binary, $ b,y \in \{0,1\} $. The dimension witness can be explicitly written as
\begin{equation}
\begin{aligned}
    T_2 = & p(0|00,0) + p(0|00,1) + p(0|01,0) - p(0|01,1) \\
    & - p(0|10,0) + p(0|10,1) - p(0|11,0) - p(0|11,1),
    \label{eq:pamT2}
\end{aligned}
\end{equation}
where \( p(b=0|x_0x_1, y) \) denotes the probability of obtaining outcome \( b=0 \) given preparation \( x = x_0x_1\) and measurement choice \( y \). Let \(\eta_0\) and \(\eta_1\) denote the detection efficiencies associated with the two measurement settings \(y = 0\) and \(y = 1\), respectively. By incorporating these efficiencies into the conditional probabilities, the effective dimension witness becomes
\begin{align}
    \tilde{T}_2 =\ & \eta_0 \big[ p(0|00,0) + p(0|01,0) - p(0|10,0) - p(0|11,0) \big] \nonumber \\
    & + \eta_1 \big[ p(0|00,1) - p(0|01,1) + p(0|10,1) - p(0|11,1) \big].
\end{align}
This expression assumes that non-detection events are absorbed into the output \(b=1\), so that the conditional probabilities \(p(b=0|x,y)\) effectively decrease in proportion to the corresponding detection efficiency \(\eta_y\), while \(p(b=1|x,y)\) absorbs the remaining probability \(1 - \eta_y\).

A common and instructive case to analyze is the symmetric case, in which both measurements are assumed to have equal detection efficiency, denoted by \(\eta_0 = \eta_1 = \eta\). Under this assumption, the effect of inefficiencies is to uniformly scale the ideal (lossless) value of the dimension witness \(T_2\) by the factor \(\eta\), resulting in a modified witness value given by \(\tilde{T}_2 = \eta T_2\). In the ideal case, the maximal quantum value of this witness is \(2\sqrt{2}\), so with symmetric inefficiencies the maximum achievable value becomes \(2\sqrt{2}\eta\). To observe a violation of the classical dimension witness bound, which is 2, the detection efficiency \(\eta\) must exceed the threshold of \(1/\sqrt{2}\), which is approximately 0.707.

In contrast, when the detection efficiencies are asymmetric (\(\eta_0 \neq \eta_1\)), the situation changes significantly. For example, one may consider the case in which one measurement basis is perfectly efficient, say \(\eta_0 = 1\), while the other has efficiency \(\eta_1 < 1\). As shown in Ref.~\cite{li2015}, even with arbitrarily low \(\eta_1 > 0\), it is still possible to achieve violations of the classical bound. This demonstrates that dimension witness inequalities can be more robust to losses when the asymmetry in the measurement setup is properly exploited.

\subsubsection{Modeling Non-Detection Events as an Additional Output}
In this approach, losses are modeled by explicitly including non-detection events as an additional outcome in the measurement process. More precisely, if the original measurement yields outcomes \(b \in \{0, \ldots, K{-}1\}\), then in the lossy scenario we consider an extended set of outcomes \(b \in \{0, \ldots, K{-}1, \varnothing\}\), where \(\varnothing\) denotes the no-click or non-detection event. Under the presence of detection inefficiencies, the observed distribution becomes
\begin{equation}
    \tilde{p}(b|x,y) =
\begin{cases}
\eta_y\, p(b|x,y), & b \in \{0, \ldots, K{-}1\}, \\
1 - \eta_y, & b = \varnothing,
\end{cases}
\end{equation}
where \(\eta_y\) is the detection efficiency associated with measurement setting \(y\). By modeling the lossy measurements as effectively mixing the ideal measurement statistics with a fixed no-detection outcome, the observed probabilities can be expressed as a convex combination:
\begin{equation}
    \tilde{p}(b|x,y) = \eta_y\, p(b|x,y) + (1 - \eta_y) p(\varnothing|x,y),
\end{equation}
where \(p(\varnothing|x,y) = \delta_{b,\varnothing}\) assigns full weight to the no-click event. In the ideal lossless scenario, the probability of the no-detection event \(p(\varnothing|x,y)\) is zero for all inputs and measurement settings, as such events do not occur. Consequently, for every valid outcome \(b \neq \varnothing\), the observed probabilities are scaled by the detection efficiency \(\eta_y\), while the probability of the no-detection outcome \(b = \varnothing\) equals the loss probability \(1 - \eta_y\).

Considering a linear dimension witness \(W(\mathbf{\tilde{p}})\) acting on the observed probabilities \(\mathbf{\tilde{p}} = \{\tilde{p}(b|x,y)\}_{b,x,y}\) as in Eq.~\eqref{eq:pam_tildeW}, one can normalize the witness such that the no-detection event contributes zero to its value, i.e., \(c_{\varnothing|x,y} = 0\) for all \(x,y\). This normalization is possible because dimension witnesses are defined up to an additive constant and an overall scaling factor without changing their operational meaning. Since the no-detection outcome represents loss rather than a meaningful measurement result, assigning zero weight to this outcome ensures that the witness value reflects only the detected events. This implies that the dimension witness value under losses scales linearly with the detection efficiency:
\begin{equation}
    W(\mathbf{\tilde{p}}) = \sum_{\substack{b \neq \varnothing \\ x,y}} c_{b|x,y} \, \eta_y \, p(b|x,y) = \eta\, W(\mathbf{p}),
\end{equation}
where, for simplicity, we assume uniform efficiency \(\eta_y = \eta\) across all measurement settings. Here, \(W(\mathbf{p})\) denotes the ideal (lossless) dimension witness value, defined in Eq.~\eqref{eq:PAM_W}. This linear scaling relation allows us to determine the minimal detection efficiency \(\eta\) required to observe a violation of the classical dimension witness bound. Specifically, if the maximum ideal quantum value of the witness is \(W_{\max}\) and the classical bound is \(W_C\), then the threshold efficiency satisfies
\begin{equation}
    \eta > \frac{W_C}{W_{\max}}.
\end{equation}
Above this threshold, the observed correlations cannot be explained by classical systems of the assumed dimension, thereby certifying either a higher dimension or the quantum nature of the source.

We illustrate an example from Ref.~\cite{dallarno2012}, which employs a generalized version of \(S_3\) (see Eq.~\eqref{eq: pam_s3}) originally introduced in Ref.~\cite{gallego2010}. The scenario is defined by the preparations \(x \in \{0, \dots, d\}\), measurements \(y \in \{0, \dots, d{-}1\}\), and three possible outcomes \(b \in \{0,1,2\}\). The corresponding generalized dimension witness is given by
\begin{equation}
    I_{d+1} = -\sum_{y=0}^{d-1} p(0|0,y) + \sum_{x=1}^{d} \sum_{y=0}^{d-x} \alpha_{xy} \, p(0|x,y),
\end{equation}
where the coefficients \(\alpha_{xy}\) are defined as
\begin{equation}
    \alpha_{xy} =
    \begin{cases}
        -1, & \text{if } x + y \leq d - 1, \\
        +1, & \text{otherwise}.
    \end{cases}
\end{equation}
For \(d = 2\), we recover a particular instance of \(S_3\). Additionally, due to the normalization of probabilities, the witness \( I_{d+1} \) effectively depends on only two out of the three possible outcomes, making it suitable for the model of efficiencies we discussed here. 

We now summarize the key results obtained in Ref.~\cite{dallarno2012}, which characterize the robustness of this witness under detection inefficiencies. In particular, the maximal quantum value of the dimension witness \(I_{d+1}\), denoted \(I^*_{d+1}\), is bounded as
\begin{equation}
    d - 2 + \sqrt{2} \;\leq\; I^*_{d+1} \;\leq\; d.
\end{equation}
The lower bound is obtained by recursively extending the optimal quantum strategy known for \(d=2\), where \(I^*_3 = \sqrt{2}\). The upper bound reflects the fact that \(d+1\) quantum states cannot be perfectly distinguished in a \(d\)-dimensional Hilbert space. These bounds on \(I^*_{d+1}\) allow us to identify two key detection efficiency thresholds, \(\eta_{\mathrm{qc}}\) and \(\eta_{\mathrm{dim}}\), which will be detailed in what follows.

The first threshold, \(\eta_{\mathrm{qc}}\), defines the minimal detection efficiency required such that any observed violation above the classical bound certifies the quantum nature of a system with dimension \(d\). This threshold is given by
\begin{equation}
    \eta_{\mathrm{qc}} := \frac{d - 1}{I^*_{d+1}},
\end{equation}
where \(d - 1\) is the maximum value attainable by classical states~\cite{gallego2010}. The threshold satisfies the bounds
\begin{equation}
    \frac{d - 1}{d} \;\leq\; \eta_{\mathrm{qc}} \;\leq\; \frac{d - 1}{d - 2 + \sqrt{2}}.
\end{equation}
For \(d=2\), \(\eta_{\mathrm{qc}}\) equals exactly \(1/\sqrt{2} \approx 0.707\). Moreover, as \(d \to \infty\), \(\eta_{\mathrm{qc}}\) approaches 1, which means that the experimental demands required to certify quantumness increase with the dimension of the system.

The second threshold, \(\eta_{\mathrm{dim}}\), corresponds to the minimal detection efficiency necessary to certify that the system has dimension at least \(d+1\). It is defined as
\begin{equation}
    \eta_{\mathrm{dim}} := \frac{I^*_{d+1}}{d},
\end{equation}
and admits the lower bound
\begin{equation}
    \eta_{\mathrm{dim}} \;\geq\; 1 - \frac{2 - \sqrt{2}}{d}.
\end{equation}
For \(d=2\), \(\eta_{\mathrm{dim}}\) is approximately \(0.707\), and similarly to \(\eta_{\mathrm{qc}}\), it tends to 1 as \(d \to \infty\).

The detection efficiency thresholds \(\eta_{\mathrm{qc}}\) and \(\eta_{\mathrm{dim}}\) can be evaluated numerically using optimization methods described in~\cite{dallarno2012}. Notably, the threshold \(\eta_{\mathrm{dim}}\), which certifies the system’s dimension, grows faster than \(\eta_{\mathrm{qc}}\), which certifies the quantum nature of the system. This behavior indicates that for a fixed dimension, distinguishing quantum from classical behavior is generally more robust to detection inefficiencies than certifying a minimal system dimension.

\subsubsection{Modeling Inefficiencies in the States}
\label{sec: effreq_for_dw}
Inefficiencies in the preparation and transmission of quantum states can be modeled using trace-preserving quantum channels. In Sec.~\ref{sec: eff_amplitude} we introduced the amplitude damping channel, which effectively describes losses due to energy dissipation during transmission. This channel models the reduction in photon number and is commonly used in optical implementations. In addition to amplitude damping, noise can also be modeled by the depolarizing channel, which captures uniform randomization of the quantum state. The depolarizing channel acts as
\begin{equation}
    \Phi_q^{\mathrm{dep}}(\rho) = (1 - q) \rho + q \frac{\mathbb{I}}{d},
    \label{eq: pam_depolarizing}
\end{equation}
where \( q \in [0,1] \) quantifies the depolarizing strength and \( d \) is the Hilbert space dimension.

In this section, we apply noise channels directly to the prepared quantum states to analyze their impact on specific dimension witnesses. This approach allows us to study the robustness of the witnesses under realistic noise conditions while keeping the measurement settings fixed. Specifically, we consider a quantum noise channel \(\Phi\), represented by a completely positive trace-preserving map, acting on the prepared states \(\{\rho_x\}\), resulting in noisy states \(\rho_x^{\mathrm{noisy}} = \Phi(\rho_x)\). Accordingly, the conditional probabilities used to evaluate the witness are modified to
\begin{equation}
    \tilde{p}(b|x,y) = \operatorname{Tr}\left(M_{b|y} \rho_x^{\mathrm{noisy}}\right) = \operatorname{Tr}\left(M_{b|y} \Phi(\rho_x)\right).
\end{equation}
Since the dimension of the states is typically known in semi-device-independent scenarios, we can parametrize both the measurements and the states to optimize the noisy behavior \(\mathbf{\tilde{p}} = \{\tilde{p}(b|x,y)\}_{b,x,y}\) by maximizing the dimension witness value \(W(\mathbf{\tilde{p}})\), as defined in Eq.~\eqref{eq:PAM_W}.

To illustrate this optimization procedure and as a  new contribution to the literature, we focus on the dimension witness \(S_3\) defined in Eq.~\eqref{eq: pam_s3}, considering a transmitted state of dimension 2. Since the system is a qubit, both the prepared states and binary-outcome measurements can be represented using Bloch vectors. Any state \(\rho_x\) and POVM element \(M_{0|y}\) can be written as
\begin{equation}
    \rho_x = \frac{1}{2}\left( \mathbb{I} + \vec{r}_x \cdot \vec{\sigma} \right), \quad
    M_{0|y} = \frac{1}{2}\left( \mathbb{I} + \vec{m}_y \cdot \vec{\sigma} \right),
\end{equation}
where \(\vec{r}_x, \vec{m}_y \in \mathbb{R}^3\) with \(\|\vec{r}_x\| \leq 1\), \(\|\vec{m}_y\| \leq 1\), and \(\vec{\sigma} = (\sigma_x, \sigma_y, \sigma_z)\) is the vector of Pauli matrices. The complementary operator is \(M_{1|y} = \mathbb{I} - M_{0|y}\). The goal is then to optimize over the Bloch vectors \(\{\vec{r}_x\}\) and measurement directions \(\{\vec{m}_y\}\) to maximize the noisy witness value \(S_3(\mathbf{\tilde{p}})\), thereby assessing the robustness of the witness against the given noise channel. Hence, the optimization problem reads
\begin{equation}
\begin{aligned}
    \max_{\{\vec{r}_x, \vec{m}_y\}} \quad & S_3(\mathbf{\tilde{p}}) \\
    \text{s.t.} \quad & \tilde{p}(b|x,y) = \operatorname{Tr}\left[ M_{b|y} \, \Phi(\rho_x) \right], \quad \forall b,x,y.
\end{aligned}
\end{equation}

\begin{figure}[!h]
    \centering
    \includegraphics[scale=0.43]{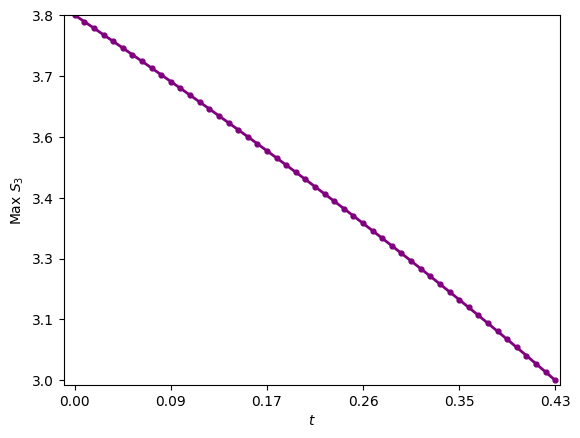}
    \caption{\justifying
    \textbf{Maximum values of the dimension witness \( S_3 \) as a function of the amplitude damping parameter \( t \).} The curve illustrates the robustness of the witness under amplitude-damping noise applied to the prepared states. The critical value \( t_c \approx 0.433 \) marks the threshold beyond which the witness no longer violates the classical bound \( S_3 = 3 \). Data points correspond to numerically optimized values for each \( t \).
    }
    \label{fig:amplitude_damping}
\end{figure}

For the amplitude damping channel \(\Phi_t^{\mathrm{AD}}\) acting on a qubit (see Eq.~\eqref{AD}), defined by the Kraus operators
\begin{equation}
    K_0 = \begin{pmatrix} 1 & 0 \\ 0 & \sqrt{1 - t} \end{pmatrix}, \quad
    K_1 = \begin{pmatrix} 0 & \sqrt{t} \\ 0 & 0 \end{pmatrix},
\end{equation}
the noisy states are
\begin{equation}
    \rho_x^{\mathrm{noisy}} = \Phi_t^{\mathrm{AD}}(\rho_x) = K_0 \rho_x K_0^\dagger + K_1 \rho_x K_1^\dagger.
\end{equation}
Solving the optimization problem for each noise parameter \(t\) yields the maximal values of the dimension witness \(S_3\) under noise, shown in Fig.~\ref{fig:amplitude_damping}. As \(t\) increases, decoherence reduces the witness value. The figure displays this degradation and identifies a critical threshold \(t_c \approx 0.433\) beyond which the classical bound \(S_3 = 3\) is no longer violated, indicating loss of dimension certification.

\begin{figure}[!h]
    \centering
    \includegraphics[scale=0.43]{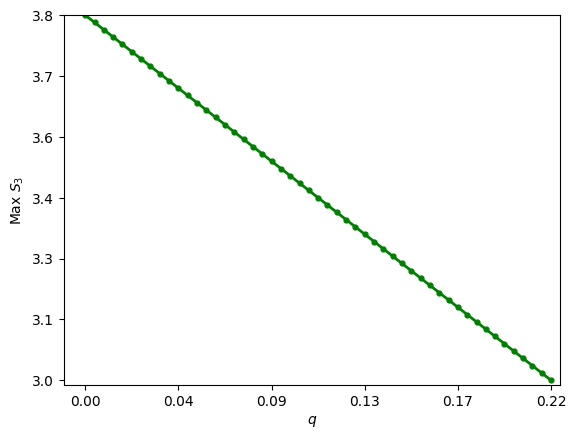}
    \caption{\justifying
    \textbf{Maximum values of the dimension witness \( S_3 \) as a function of the depolarizing parameter \( q \).} The curve demonstrates the robustness of the witness under depolarizing noise applied to the prepared states. The critical value \( q_c \approx 0.216 \) indicates the noise level above which the witness no longer exceeds the classical bound \( S_3 = 3 \). Data points represent numerically optimized values for each \( q \).
    }
    \label{fig:depolarizing}
\end{figure}

For the depolarizing channel \(\Phi_q^{\mathrm{dep}}\) acting on a qubit, defined by Eq.~\eqref{eq: pam_depolarizing} with \(d = 2\), the noisy states are
\begin{equation}
    \rho_x^{\mathrm{noisy}} = \Phi_q^{\mathrm{dep}}(\rho_x) = (1 - q) \rho_x + q \frac{\mathbb{I}}{2}.
\end{equation}
Solving the optimization problem for each noise parameter \(q\) yields the maximal values of the dimension witness \(S_3\) under noise, shown in Fig.~\ref{fig:depolarizing}. As \(q\) increases, the prepared states become more mixed, decreasing the witness value. The figure illustrates this decay and identifies a critical threshold \(q_c \approx 0.216\).

In Ref.~\cite{li2015}, the authors study the effect of depolarizing noise on the dimension witness \(T_2\) in Eq.~\eqref{eq:pamT2}, under the assumption of perfect measurements and state preparations affected by white noise. They show that the observed value of the witness scales linearly with the noise parameter as \(T_{\mathrm{practical}} = (1 - q){T}_2\), where \({T}_2\) is the ideal (noise-free) value. From this, they derive a critical noise threshold \(q_c = 1 - 2/T_2\) above which no violation of the classical bound \(T_2 \leq 2\) is possible. For the optimal quantum violation \(T_2 = 2\sqrt{2}\), this yields \(q_c = 1 - \sqrt{2}/2 \approx 0.293\). The authors also analyze how this threshold changes in the presence of detection inefficiencies, providing numerical results for both symmetric and asymmetric detection scenarios using the absorption model described in Sec.~\ref{sec: pam_absorption}. 

An amplitude-damping model is also used in Ref.~\cite{alves2025}, which considers a discretized continuous-variable optical implementation and studies losses in PAM scenarios with three and four preparations and two or three measurement settings. Measurements are implemented via binned homodyne detection or via displacement followed by on/off photodetection, with losses modeled through amplitude damping acting only on the displacement-based measurements, while homodyne detection is treated as ideal.

\section{The Bilocality Scenario}
\label{sec:The Bilocality Scenario}

\begin{figure}[h!]
\centering
\begin{tikzpicture}[
    node distance=1.5cm,
    abc circle node/.style={circle, draw=blue, fill=blue!30, minimum size=1cm, inner sep=0pt, font=\large\itshape}, 
    other circle node/.style={circle, draw=green, fill=green!30, minimum size=1cm, inner sep=0pt, font=\large\itshape}, 
    every rectangle node/.style={rectangle, draw=orange, fill=orange!30, minimum size=1cm, inner sep=0pt, font=\large\bfseries}, 
    ->, >={Latex}
]

\node[abc circle node] (A) at (0,0) {A};
\node[abc circle node] (B) [right of=A, xshift=2cm] {B};
\node[abc circle node] (C) [right of=B, xshift=2cm] {C};
\node[every rectangle node] (Lambda1) at ($(A)!0.5!(B)+(0,1)$) {$\lambda_1$};
\node[every rectangle node] (Lambda2) at ($(B)!0.5!(C)+(0,1)$) {$\lambda_2$};
\node[other circle node] (X) [above of=A, yshift=0.5cm] {X};
\node[other circle node] (Y) [above of=B, yshift=0.5cm] {Y};
\node[other circle node] (Z) [above of=C, yshift=0.5cm] {Z};

\draw[->] (X) -- (A);             
\draw[->] (Y) -- (B);             
\draw[->] (Lambda1) -- (A);       
\draw[->] (Lambda1) -- (B);       
\draw[->] (Z) -- (C);             
\draw[->] (Lambda2) -- (C);       
\draw[->] (Lambda2) -- (B);       
\end{tikzpicture}
\caption{\textbf{Directed acyclic graph representing a bilocality scenario.}}
\label{fig:bilocal}
\end{figure}
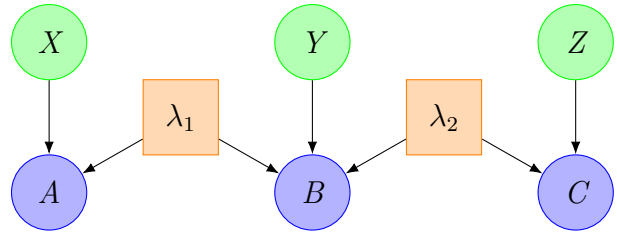

In multipartite Bell scenarios, it is assumed that the correlations between the measurement outcomes of all distant parties share a common hidden variable \cite{Chaves2017causalhierarchyof}. This may happen, for instance, if all parties receive particles that were produced by a single source. However, in an experiment that involves multiple independent sources, one needs to consider more general causal structures than those underlying Bell scenarios \cite{Tavakoli_2022, chaves2015unifying}. A simple example is the scenario underlying entanglement swapping experiments \cite{pan1998experimental}, in which there are two sources of particles and three distant parties  - let us call them Alice, Bob and Charlie. In this scenario, source 1 sends an entangled pair to Alice and Bob, and source 2 sends another pair to Bob and Charlie.
Since it is assumed that the sources are initially independent, the classical description of the correlations between measurement outcomes must be given in terms of independent hidden variables $\lambda_1$ and $\lambda_2$. If Alice, Bob and Charlie's measurement choices are denoted by $x,y$ and $z$, and their outcomes by $a,b$ and $c$, respectively, then any probabilistic model compatible with local-realism, supplemented by the additional assumption of independence of the sources, must have the form
\begin{multline}
    p_{B}(a,b,c|x,y,z) = \int_{\Lambda_1} d\lambda_1\int_{\Lambda_2} d\lambda_2\, q_1(\lambda_1)q_2(\lambda_2)\\\times p(a|x,\lambda_1)p(b|y,\lambda_1,\lambda_2)p(c|z,\lambda_2).
    \label{bilocal model}
\end{multline}
with $q_1(\lambda_1)$ and $q_2(\lambda_2)$ being the probability densities of the corresponding independent hidden variables. A probabilistic model of the form (\ref{bilocal model}) is called a bilocal hidden variable model (BLHV)\cite{branciardcharacterizing, branciardbilocal} with several experiments performed to this date \cite{carvacho2017experimental,saunders2017experimental,sun2019experimental,andreoli2017experimental}. To distinguish a tripartite scenario with two independent hidden variables from the usual tripartite Bell scenario, the former is called a bilocality scenario, whose corresponding causal structure is shown in Fig.~\ref{fig:bilocal}. 

A bilocality scenario is fixed by specifying the number of measurement choices and outcomes for each party. Here, we focus on the scenario where Alice and Charlie measure two possible dichotomic observables each $(A_0,A_1,C_0,C_1)$, while Bob performs a fixed measurement with four possible outcomes $b\equiv (b^0,b^1)\in\{(0,0),(0,1),(1,0),(1,1)\}$, which, in the case of entanglement swapping, is usually assumed to be a Bell State Measurement (BSM) of the qubits that he receives from each of the sources \cite{zukowski1993event}. Since Bob has one input and four outputs, this is called the $(1,4)$ bilocality scenario. In this scenario, any bilocal hidden variable model has the form
\begin{multline}
    p_B^{14}(a,b^0,b^1,c|x,z) = \int_{\Lambda_1} d\lambda_1\int_{\Lambda_2} d\lambda_2\, q_1(\lambda_1)q_2(\lambda_2)\\\times p(a|x,\lambda_1)p(b^0,b^1|\lambda_1,\lambda_2)p(c|z,\lambda_2).
    \label{bilocal 14 model}
\end{multline}

 For any correlation compatible with the Eq.(\ref{bilocal 14 model}), the so-called IJ inequality has been proven to hold \cite{branciardbilocal}:
\begin{equation}
\mathcal{B}\equiv\sqrt{\abs{I^{14}}}+ \sqrt{\abs{J^{14}}}\leq 1,
\label{IJ inequality}
\end{equation}
with $I^{14}$ and $J^{14}$ defined by $ I^{14} = \frac{1}{4}\sum_{x,z=0}^1\expval{A_xB^0C_z}$, $ J^{14}= \frac{1}{4}\sum_{x,z=0}^1(-1)^{x+z}\expval{A_xB^1C_z}$, where $\expval{A_x B^y C_z} = \sum_{a,b^0,b^1,c}(-1)^{a+b^y+c}p^{14}(a,b^0,b^1,c|x,z)$.

In contrast, if Alice and Bob share a quantum state $\rho_1$ produced by the source 1, while Bob and Charlie share another state $\rho_2$ produced by the source 2, the assumption of independence of the sources requires that the density operator of the total system be a product state $\rho_{ABC}= \rho_1\otimes \rho_2$. Hence, in this case, the Born rule may be expressed as what is called a biquantum model:
\begin{equation}
    p^{14}_{QM}(a,b^0,b^1,c|x,z) = \mathrm{Tr}[M^A_{a|x}\otimes M^B_{b^0,b^1}\otimes M^C_{c|z}\,\rho_1\otimes \rho_2],
\end{equation}
where $M^A_{a|x}, M^B_{b^0,b^1}$ and $M^C_{c|z}$ are the measurement operators of Alice, Bob and Charlie, respectively. If both $\rho_1$ and $\rho_2$ are maximally entangled two-qubit states and Bob performs a BSM, then, with suitable local measurements of Alice and Charlie, values of $\mathcal{B}$ as large as $\mathcal{B}= \sqrt{2}$ can be attained, violating the IJ inequality (\ref{IJ inequality}), and thus proving that no bilocal hidden variable model can reproduce such (bi)quantum correlations. 

\subsection{Efficiency Requirements in the Bilocal Scenario}
We now describe the modeling of limited detection efficiencies in the $(1,4)$ bilocality scenario. This is usually done under the simplifying assumption that, in every run of the experiment, Bob successfully performs a BSM - that is, Bob's detection device is treated as perfect. This is a valid simplification if the focus is to understand how inefficient the measurement devices of Alice and Charlie can be and still produce correlations that cannot be explained by a bilocal model. Following Ref. \cite{boundingthesets}, we thus consider that for all measurements performed by Alice, there is a probability $1-\eta_1$ to observe a third (lossy) outcome, where $\eta_1$ denotes the efficiency of her detector. Similarly, if the efficiency of Charlie's detector is $\eta_2$, then he observes a lossy outcome with probability $1-\eta_2$. In this case, the probability when both Alice's and Charlie's detectors click is 
\begin{equation}
    p_{\eta_1 \eta_2}^{14}(a,b^0,b^1,c|x,z) = \eta_1\eta_2 p_{QM}^{14}(a,b^0,b^1,c|x,z).
    \label{P14real}
\end{equation}
This implies that the value of expression (\ref{IJ inequality}), when these detection inefficiencies are taken into account, is given by
\begin{eqnarray}
    \mathcal{B}_{real}(\eta_1,\eta_2) = \sqrt{\eta_1\eta_2}\mathcal{B}_{ideal},
\end{eqnarray}
with $\mathcal{B}_{ideal}$ being the value of the IJ expression for perfect detection efficiencies. 

In order for the distribution (\ref{P14real}) to violate the IJ inequality, one must have $ \mathcal{B}_{real}(\eta_1,\eta_2)>1$. Since the largest value of $\mathcal{B}_{ideal}$ that can be attained quantum-mechanically is $\mathcal{B}_{ideal}=\sqrt{2}$, one can conclude that for symmetric inefficiencies $\eta_1=\eta_2=\eta$, the  IJ inequality is violated only for $ \eta>\frac{1}{\sqrt{2}} \approx 0.7071$. This value is an upper bound for the \textit{bilocal detection efficiency threshold}, $\eta_{biloc}$, defined as the largest detection efficiency for which the correlation $P^{14}_{real}$ admits a
bilocal decomposition. In fact, as reported in Ref. \cite{boundingthesets}, if the state between Alice and Bob is fixed to be maximally entangled, correlations
without bilocal models can be established for symmetric efficiencies
above $0.6085$. Additionally, if one considers that the two sources $i\in\{1,2\}$ produce pure entangled states $\ket{\psi_i}=\cos\theta_i\ket{00}+\sin\theta_i\ket{11}$, then assuming $\theta_1=\theta_2$ and varying the entanglement of $\ket{\psi_1}$, nonbilocal correlations can be generated for efficiencies higher than $0.5291$, which is far below the lowest efficiency required to violate the IJ inequality. This gap illustrates the fact that inequality (\ref{IJ inequality}) is a necessary, but not sufficient, condition for bilocality, as discussed in detail in Ref. \cite{polynomialBell}.

A second possibility is for Alice's detection device to be perfect $(\eta_1=1)$, in which case the inequality (\ref{IJ inequality}) is violated for $\eta_2>0.5$. On the other hand, Ref. \cite{boundingthesets} reports that nonbilocal correlations can be established even for detection efficiencies as low as $\eta_2 = 0.00001$. Again, this discrepancy between the efficiency required to violate the IJ inequality and the detection threshold is a consequence of the fact that the IJ inequality alone does not provide a complete description of bilocal correlations. In fact, other inequivalent bilocality inequalities for the $(1,4)$ scenario have been introduced, such as in Ref. \cite{tavakoli2021bilocal}. 

Yet another important point of discussion is that the required detection efficiencies to certify nonbilocality can be smaller than those needed to discard local models in a standard Bell scenario, such as the ones reported in Ref. \cite{massar2003violation}. For instance, it is well known that if Alice and Charlie run a standard bipartite Bell test, possibly conditioned on Bob's measurement output, then, for symmetric detection efficiencies $\eta_1=\eta_2=\eta$, a local model for the correlations always exists
if $\eta\leq2/3\approx0.667$, and in case one device is perfect, say $\eta_1=1$, it is not possible to observe nonlocal correlations for $\eta_2< 0.5$ \cite{boundingthesets} . A comparison with the previously mentioned results for minimal bilocal detection efficiency reported in this same reference reveals a considerable gap between the nonlocality detection threshold and the corresponding bilocality detection threshold. Since bilocal correlations are more constrained than local models due to the additional assumption of independence of the sources, this conclusion is perhaps unsurprising. However, it allows one to take advantage of the network structure of a bilocality scenario to lower the detection efficiency requirements for certifying nonclassical correlations, and hence infer the presence of entanglement.

\begin{widetext}
\centering
\small
\renewcommand{\arraystretch}{1.05}

\captionof{table}{Overview of representative detection-efficiency thresholds in Bell, instrumental, PAM, and bilocal scenarios. Rows detail benchmarks, non-detection models, critical efficiencies ($\eta_{\rm crit}$), and the specific contributions of this review. Results marked with $\dagger$ represent new findings derived in this paper.}
\label{tab:scenarios_justified}

\begin{tabular*}{1\textwidth}{@{\extracolsep{\fill}} llllll @{}}
\hline
\textbf{Scenario} & 
\textbf{Setup} & 
\textbf{ Model} & 
\textbf{$\eta_{\rm crit}$} & 
\textbf{Contribution} & 
\textbf{References} \\ \hline

Bell & 
\begin{minipage}[t]{0.13\textwidth}\sloppy CHSH $(2,2,2)$; CH/CHSH forms; $I_{3322}$; multipartite Mermin/Svetlichny\end{minipage} & 
\begin{minipage}[t]{0.11\textwidth}\sloppy Extra-outcome: explicit $\varnothing$; Absorption: no-click $\to$ output; Hybrid\end{minipage} & 
\begin{minipage}[t]{0.11\textwidth}\sloppy Sym: $2/3 \approx 0.667$ (CH); $0.84$ (others); Asym: $0.43$; Mermin: $0.81$\end{minipage} & 
\begin{minipage}[t]{0.26\textwidth}\sloppy Unified analysis of detection-efficiency thresholds in Bell tests (Sec.~\ref{sec:The Bell scenario}) for CH/CHSH, $I_{3322}$ and multipartite Mermin/Svetlichny inequalities under extra-outcome, absorption and hybrid loss models, with optimized symmetric and asymmetric efficiency curves.\end{minipage} & 
\begin{minipage}[t]{0.08\textwidth}\cite{Clauser1969,Clauser.Horne,Eberhard93,Larsson_Bell,BrunnerREVIEW,Cabello2009,BrunnerI3322,CabelloLarsson,Massar_Pironio_2003,Garbarino2010,Cao2016,Mermin_ineqs,Svetlichny_ineqs,Cabello_Mermin,Pal_Vertesi_GHZ,Pal_Vertesi_W,M_measurements_1,M_measurements_2,M_measurements_3,efficiencies_Subhendu_2024,Gebhart_2022_GMN,Polonia_Tamas_Multipartite,Simmetric_Multipartite_designolle,Chaves2017causalhierarchyof,polynomialBell}\end{minipage} \\[10em] \hline

Instrumental$^\dagger$ & 
\begin{minipage}[t]{0.13\textwidth}\sloppy $I_{222}$, $I_{223}$, $I_{233}$ with observational and interventional probabilities\end{minipage} & 
\begin{minipage}[t]{0.11\textwidth}\sloppy Extra-outcome: explicit $\varnothing$ on nodes; Absorption: no-click $\to$ fixed; Hybrid: mixed approaches\end{minipage} & 
\begin{minipage}[t]{0.11\textwidth}\sloppy $I_{222}$: Sym $0.67$, Asym $0.50$; $I_{223}$: Sym $0.90$, Asym $0.51$; $I_{233}$: Sym $0.9052$, Asym $0.51$ or $0.8750$\end{minipage} & 
\begin{minipage}[t]{0.26\textwidth}\sloppy $^\dagger$First systematic analysis of instrumental inequalities under different loss models (Sec.~\ref{sec:The Instrumental Scenario}). Note that Fig.~(\ref{fig4}) was first obtained in \cite{ObsInt_Bell}.\end{minipage} & 
\begin{minipage}[t]{0.08\textwidth}\cite{pearl1995causal,Pearl2009causality,pearl2013,bonet_inequalities,inst_inequalities_x,van2019quantum,chaves2018quantum,chaves2021causal,Instrumental_Miklin,ObsInt_Bell,GMR_QCI,PhysRevA.104.L010201,chaves2015unifying,Chaves2017causalhierarchyof,ObsInt_Bell}\end{minipage} \\[10em] \hline

Prepare-and-measure~$^\dagger$ & 
\begin{minipage}[t]{0.13\textwidth}\sloppy Dimension witness $I_{d+1}$ for bounded dimension $d$; qubit case $d=2$; quantumness vs.\ dimension certification\end{minipage} & 
\begin{minipage}[t]{0.11\textwidth}\sloppy Extra-outcome: explicit $\varnothing$ on measurement device; combined source-detector efficiency\end{minipage} & 
\begin{minipage}[t]{0.11\textwidth}\sloppy $\eta_{\rm qc}$ (quantumness); $\eta_{\rm dim} \approx 0.707$ ($d=2$); both increase with $d \to 1$ as $d \to \infty$; $\eta_{\rm qc} < \eta_{\rm dim}$\end{minipage} & 
\begin{minipage}[t]{0.26\textwidth}\sloppy Unified treatment of dimension witnesses under losses, including $^\dagger$robustness analysis of the qubit $S_3$ witness under amplitude-damping and depolarizing noise (Figs.~\ref{fig:amplitude_damping}–\ref{fig:depolarizing}), clarifying how inefficiencies limit semi-device-independent quantum resource certification (Sec.~\ref{sec:The Prepare-and-Measure Scenario}).\end{minipage} & 
\begin{minipage}[t]{0.08\textwidth}\cite{gallego2010,brunner2008,bowles2014,Gallego_NL_2012,navascues2015,dallarno2012,davide2020,pawlowski2010,pawlowski2011,li2011,li2012,li2015,DallArno2015,Mironowicz2021,hendrych2012,ahrens2012,ahrens2014,Ambainis2008,bendersky2016algorithmic,zamora2025,alves2025}\end{minipage} \\[10em] \hline

Bilocality & 
\begin{minipage}[t]{0.13\textwidth}\sloppy Tripartite network: Alice, Charlie $(2,2)$; Bob 4-outcome Bell measurement; IJ inequality $(1,4)$; improved witnesses\end{minipage} & 
\begin{minipage}[t]{0.11\textwidth}\sloppy Extra-outcome on end nodes Alice, Charlie; Bob ideal; optimized measurements for improved witnesses\end{minipage} & 
\begin{minipage}[t]{0.11\textwidth}\sloppy IJ-type: Sym $\eta > 1/\sqrt{2} \approx 0.7071$; Improved: $\eta \gtrsim 0.6085$ (below Bell $2/3$ threshold)\end{minipage} & 
\begin{minipage}[t]{0.26\textwidth}\sloppy Review of bilocal detection thresholds, showing how network structure with independent sources relaxes efficiency requirements compared to standard Bell tests (Sec.~\ref{sec:The Bilocality Scenario}).\end{minipage} & 
\begin{minipage}[t]{0.08\textwidth}\cite{zukowski1993event,pan1998experimental,branciardcharacterizing,branciardbilocal,boundingthesets,Tavakoli_2022,tavakoli2021bilocal,saunders2017experimental,carvacho2017experimental,andreoli2017experimental,scheidl2010violation,ObsInt_Bell,Chaves2017causalhierarchyof}\end{minipage}\\[10em]  

\end{tabular*}
\end{widetext}
\section{Final Remarks}
\label{sec:final}

Bell’s theorem and its various generalizations to different types of quantum networks remain among the most striking manifestations of quantum non-classicality, fundamentally challenging our intuitions about cause and effect, and, more broadly, our common-sense understanding of physical reality. Beyond its foundational significance, the violation of Bell inequalities plays a pivotal role in (semi-)device-independent frameworks for quantum information processing. These frameworks enable practical quantum technologies and quantum advantages by relying solely on observed statistical correlations as a resource—correlations that guarantee security or unpredictability independently of the internal workings of the devices, with key applications in quantum communication protocols such as randomness generation, randomness certification, and quantum key distribution.

However, the practical realization of such protocols entails overcoming substantial experimental challenges. In the context of Bell inequality violations and related applications, this primarily involves the closure of experimental loopholes. Among them, the detection loophole, stemming from imperfect detectors and inevitable losses, has long hindered the conclusive demonstration of non-classical correlations. If left unaddressed, this loophole allows local hidden variable models to reproduce quantum-like statistics, thus undermining both foundational claims and the security of quantum communication protocols. As such, understanding and quantifying the detection efficiency thresholds required for various non-classicality scenarios is not a mere academic exercise—it is a fundamental prerequisite for the reliable deployment of quantum technologies in real-world settings.

This review has offered a unified perspective on how detection efficiency requirements manifest across a range of scenarios designed to test quantum non-classicality. Starting with the Bell scenario and extending to instrumental, prepare-and-measure, and bilocality frameworks, we have highlighted how the underlying causal structure strongly influences the required detection efficiencies. In some cases, such as quantum networks with independent sources—the structure of the correlations can relax efficiency requirements, making the observation of non-classicality more experimentally feasible. Conversely, in semi-device-independent settings like dimension witnesses, detector inefficiencies can directly jeopardize the certification of quantum resources, making the establishment of stringent efficiency thresholds especially critical.

Looking ahead, this field continues to present both significant opportunities and technical challenges. On the one hand, future developments could include the design of non-classicality tests tailored to specific experimental conditions, including losses, noise, and finite statistical data. On the other hand, the growing theoretical understanding and experimental implementation of quantum networks opens exciting new avenues for leveraging complex correlation structures to reduce practical demands.

Finally, while quantum communication remains the most immediate application area, efficient and loophole-resistant non-classicality tests, especially in the device-independent regime, also hold promise for quantum computing, quantum metrology, and the study of large-scale quantum networks. These directions remain relatively unexplored, but as the field advances, so will the need for increasingly refined tools to certify and harness quantum resources under realistic constraints.

\section*{Author Contributions}

T.S.S., S.Z., M.A., and V.F.A. conducted the investigation, performed the numerical simulations, and wrote the main manuscript text. R.C. and G.V. contributed with ideas, interpretation, and critical review of the manuscript. T.S.S. contributed to the conceptualization of the project, which was supervised by R.C. All authors reviewed and approved the final manuscript. The code used to generate the results in this paper is available at
\url{https://gitlab.com/fszba/minimum-efficiencies}.

\section*{Acknowledgements}
This work was supported by the Simons Foundation (Grant Number 1023171, RC), the Brazilian National Council for Scientific and Technological Development (CNPq, Grants No.403181/2024-0 , 301687/2025-0 and 302414/2022-3), the Financiadora de Estudos e Projetos (grant 1699/24 IIF-FINEP) and the Coordenação de Aperfeiçoamento de Pessoal de Nível Superior – Brasil (CAPES) – Finance Code 001. This work has also been partially funded by the project "Comparative Analysis of P\&M Protocols for Quantum Cryptography" supported by QuIIN - Quantum Industrial Innovation, EMBRAPII CIMATEC Competence Center in Quantum Technologies, with financial resources from the PPI IoT/Manufatura 4.0 of the MCTI grant number 053/2023, signed with EMBRAPII.

\bibliography{references}

\end{document}